\documentclass[sigconf]{acmart}

\AtBeginDocument{%
  }

\copyrightyear{2026}
\acmYear{2026}
\setcopyright{cc}
\setcctype{by}
\acmConference[DIS '26]{Designing Interactive Systems Conference}{June 13--17, 2026}{Singapore, Singapore}
\acmBooktitle{Designing Interactive Systems Conference (DIS '26), June 13--17, 2026, Singapore, Singapore}
\acmISBN{979-8-4007-2563-0/2026/06}
\acmDOI{10.1145/3800645.3812971}

\newcommand{\system}[0]{AnimationDiff}
\newcommand{\hl}[1]{{\color{black}#1}}

\usepackage{makecell}

\begin{document}

\title[\system{}]{\system{}: A Visual Comparison Tool for Generated 3D Character Animations}
\settopmatter{authorsperrow=2}


\author{Ludwig Sidenmark}
\orcid{0000-0002-7965-0107}
\affiliation{%
 \institution{University of Toronto}
  \institution{Autodesk Research}
 \city{Toronto}
 \state{Ontario}
 \country{Canada}}
\email{lsidenmark@dgp.toronto.edu}

\author{Qian Zhou}
\orcid{0000-0002-7294-2561}
\affiliation{%
 \institution{Autodesk Research}
 \city{San Francisco}
 \state{California}
 \country{USA}}
\email{qian.zhou@autodesk.com}

 \author{George Fitzmaurice}
\orcid{0000-0002-2834-7757}
\affiliation{%
 \institution{Autodesk Research}
 \city{Toronto}
 \state{Ontario}
 \country{Canada}}
 \email{george.fitzmaurice@autodesk.com}

 \author{Fraser Anderson}
\orcid{0000-0003-3486-8943}
\affiliation{%
 \institution{Autodesk Research}
 \city{Toronto}
 \state{Ontario}
 \country{Canada}}
\email{fraser.anderson@autodesk.com}

\renewcommand{\shortauthors}{Sidenmark et al.}

\begin{abstract}
  Creating 3D character animations traditionally requires significant time and effort from the animator. Advancements in generative \hl{methods} now enable easy creation of multiple character animation variations for use or further editing. However, this capability introduces a new challenge in comparing character animations to select the best animation, which is challenging due to temporal misalignment and the large amount of spatial data. We present \system{}, a visual comparison tool for generated character animations. \system{} enables contextual comparisons in the intended scene and camera angle, and embedding of spatial information by combining established animation visualization techniques and easy switching between overlaid and side-by-side comparisons. \system{} also supports filtering to handle information overload, and Temporal Lenses that visualize entire animations over time for overview, alignment, and comparison.  We evaluated \system{} in a user study, showcasing its efficacy in animation comparison and providing design insights for comparing motion. 
\end{abstract}


\begin{teaserfigure}
 \centering
  \includegraphics[width=.88\textwidth]{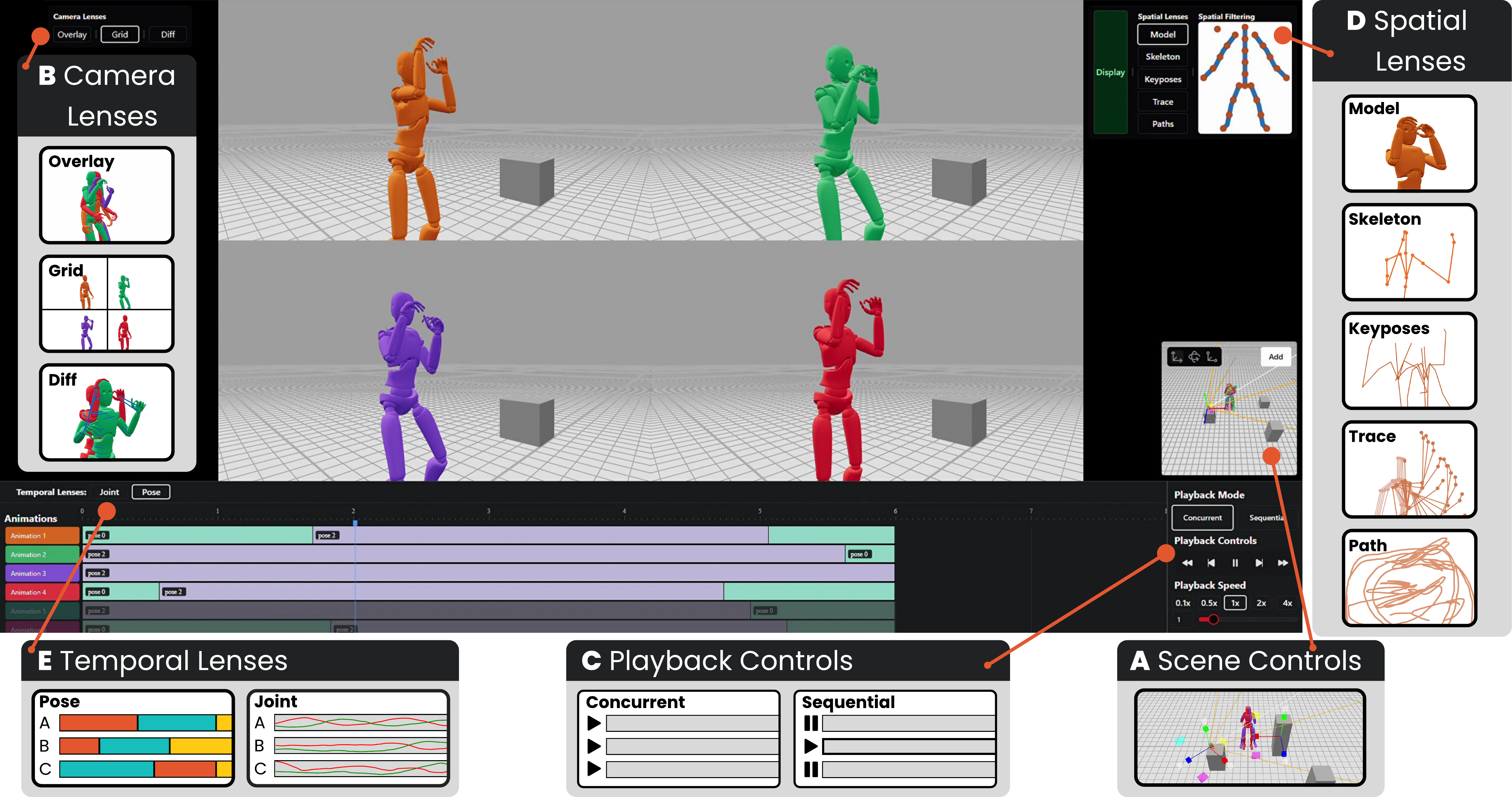}
  \caption{\system{} for evaluation and comparison of generated 3D character animations. (a) Scene controls enable users to contextualize the animations in camera and scene. (b) Camera Lenses enable users to switch between comparison modes. (c) Playback controls enable control which and how animations are played in the scene. (d) Spatial Lenses visualize and filter spatial information in the scene. (e) Temporal Lenses give an overview of animations over time for skimming and quick comparison.}
  \label{fig:teaser}
  \Description{This figure shows an overview of the system with its components highlighted. In the center are 4 animations of a character performing a boxing animation. There is a left sidebar that shows a toggle between three camera lenses (Overlay, Grid, Diff) each visualized. On the top right there is a menu showing a toggle of spatial lenses (model, skeleton, keyposes, trace, path), and a skeleton visualization which users can select to show or hide individual joints. In the bottom right, there is a thumbnail of the scene from a different camera vantage point for scene editing. Below that is a menu containing playback controls and a toggle for different playback modes (concurrent, sequential). At the bottom there is a timeline with a bar for each animation, and a needle to show the current play time. Above the bar to the left the users can toggle between different temporal bar visualizations (Pose, Joint).
}
\end{teaserfigure}

\maketitle

\section{Introduction}

3D character animation has traditionally been a resource- and time-intensive process that requires significant skill and effort from animators.  Substantial efforts have therefore been made to create animation tools such as inverse-kinematics models to reduce manual steps~\cite{Ciccone2019AnimationControl}, and new modalities~\cite{Zhou2024TimeTunnel, Fender2015CreatureTeacher} to make the creation of character animations faster and more accessible. However, character animation remains a laborious process. As such, significant creative decisions and explorations are often made early in the storyboarding phase to save cost and effort~\cite{price2015storyboarding}.

Generative artificial intelligence (GenAI) has recently seen significant advances in its capability to generate character motion using inputs such as text prompts~\cite{jiang2024motiongpt}, motion trajectories~\cite{Wan2025TLControl} or reference images~\cite{Jiang2025MotionChain}. These advancements have made it easy to quickly create a set of variations for a character animation. This capability enables animators to quickly experiment with different variations to help their creative process, but creates a new challenge in evaluating and comparing character animations to select the best one for their needs. Generated animations often differ in the timing, spatial execution, and speed of their actions, making direct comparison difficult. Visualization methods, including the display of joints and motion paths, are commonly used to represent and emphasize movement~\cite{Zhou2024TimeTunnel}. However, it remains uncertain how these methods can be applied to compare different animations, or how users might combine them to efficiently understand differences. 

Consider a set of animations depicting a boxing gesture; the gesture may begin and end at different times, involve different arms, exhibit varying ranges of motion, or include pauses of differing durations. Such discrepancies lead to temporal and spatial misalignments that require substantial manual adjustment before meaningful comparisons can be made. Direct pose-to-pose comparison is also problematic, as the corresponding poses may not be visually or functionally similar. The complexity is further compounded by the sheer volume of spatio-temporal data, which can easily overwhelm users and result in significant occlusion during visual inspection. Unlike other analysis of 3D spatio-temporal data, such as gesture elicitation or spatial recordings that primarily focus on distinct poses~\cite{Dang2021GestureMap, Reipschlager2022Avatar}, character animations demand consideration of timing, speed, transitional dynamics, and stylistic elements when assessing quality and suitability~\cite{thomas1981illusion}. Moreover, the perceived quality of an animation can vary substantially depending on its intended environment and the specific camera viewpoint, adding another layer of complexity unique to 3D animation evaluation. \hl{In this work, we explore ways to help animators compare and select the most suitable animation from a set of alternatives. While motivated by generative pipelines due to their ease of creating multiple variations, our goal is to support comparison of character animations more broadly, regardless of how they are authored.} 

To guide our research, we conducted formative interviews with five character animators to understand how they review and evaluate the quality of animations. The results of these interviews and previous work on comparing spatial and temporal data~\cite{Huh2025VideoDiff, Wang2024MotionComparator, Dang2021GestureMap} led to multiple design goals. These design goals highlight the importance of \textit{evaluating animations in their intended context} (\textbf{DG1}), giving an \textit{overview of the entire animations} for easy comparison (\textbf{DG2}), including \textit{multi-mode comparison} to support different comparison needs (\textbf{DG3}), and enabling \textit{filtering of information} to handle occlusion and information overload (\textbf{DG4}). Based on these design goals, we designed and implemented \emph{\system{}} (\autoref{fig:teaser}) \hl{for visual comparison of 3D character animations.} 

\system{} integrates candidate animations into a common environment that enables animators to add objects to the environment and adjust the camera to view and compare the animations in their intended context (\autoref{fig:teaser}a). Users can easily switch between three \textbf{Camera Lenses} (\autoref{fig:teaser}b) for different comparison modes based on established information visualization methods: 1) \emph{Overlay} for overlaid superposition comparison; 2) \emph{Grid} for side-by-side juxtaposition comparison; and 3) \emph{Diff} for explicit encoding-based comparison of animation joints. These can be combined with \emph{Concurrent} and \emph{Sequential} \textbf{Playback Modes} (\autoref{fig:teaser}c) for simultaneous and single playback, respectively, to enable direct comparison of timing and movements, and visual inspection of singular animations. To aid in spatial comparison, \system{} integrates \textbf{Spatial Lenses} (\autoref{fig:teaser}d) into the scene based on commonly used spatial visualizations in animation to highlight the character model and its skeleton and joint positions in the current frame (\emph{Skeleton}), and their movements at keyframes (\emph{Keyposes}), in a time-interval before and after the current frame (\emph{Trace}), and across the whole animation (\emph{Path}). Finally, \system{} provides two novel \textbf{Temporal Lenses} (\autoref{fig:teaser}e) that provide an overview of whole animations: 1) \emph{Joint} curves to visualize how joint positions move across the display over time; 2) \emph{Pose} segments which globally cluster poses and visualize their transitions over time to establish a common ground and enable comparison without requiring visual inspection and manual alignment of the animations. 

We evaluated \system{} in a user study with 12 participants with experience in animation creation and usage. Participants had to spatially, temporally, and aesthetically \hl{compare a set of animation variants} in a series of tasks using \system{}. The results showed that \system{} was easy to use and the integration of multi-mode system features were useful in comparing 3D character animations of various lengths and motions.

In summary, the paper presents the following contributions:
\begin{itemize}
    \item The \system{} system which integrates and visualizes spatial and temporal information, together with the scene context, to aid animation comparison.
    \item Results and associated design goals from a formative study with five animators on their animation review practices, showing the importance of in-scene evaluation and a \hl{diverse toolset} to suit animators' various evaluation needs.
   \item Results of a user study with 12 participants using \system{}, showing that visualizations spanning larger time intervals are effective for comparison. Different evaluations (spatial, temporal, and naturalness) and animation aspects (length, root motion) require different comparison tools, confirming the effectiveness of \system{}’s multi-mode design, and in-scene context enables informed decision-making. Our results provide insights for future work on animation creation tools and AI integration into animator workflows.
\end{itemize}
\section{Related Work}
This work builds on previous work on visual comparison and sensemaking, visualizing temporal and spatio-temporal data, and exploring generative data sets.

\subsection{Visual Comparison and Sensemaking}
Visual comparison is a fundamental task in creative and knowledge-intensive tasks for sensemaking and understanding the design space~\cite{gebreegziabher2024supporting}, ultimately aiding the creator to the right design~\cite{Tohidi20016RightDesign}. Three main approaches for visual comparison have been defined: superposition, juxtaposition, and explicit encoding~\cite{Gleicher2011VisualComparison}. \emph{Superposition} (or overlay) visualization designs show multiple objects in the same frame of reference (i.e., on top of each other). \emph{Juxtaposition} approaches place objects separately in place and relies on the viewer's memory to make connections between objects. Finally, \emph{explicit encoding} approaches calculate the relationship between objects, for example the difference between objects, and provide visual encoding of the relationships. These techniques are well established in information visualization and has been used for a wide range of applications in 2D and 3D~\cite{Gleicher2011VisualComparison, Kim2017Survey}. We use these techniques in the design of \system{} Camera Lenses to support different types of 3D character animation comparison.

\subsection{Evaluating Generated and Temporal Data}
To support the capabilities of generative models and encourage divergent thinking in creative processes~\cite{Dow2011Divergent, Suh2024Luminate} recent works have explored novel ways for users to compare and contrast multiple generations for image generation~\cite{Brade2023Promptify, Benharrak2025HistoryPalette}, prototyping~\cite{Zhang2025FusionProtor}, \hl{and generative design~\cite{Pandey2023Juxtaform, Matejka2018DreamLens}.} These works have been shown to aid efficiency and creativity when working with generative models by providing multiple alternatives, but are limited to static content. 

The temporal dimension of content such as videos and animations provides additional challenges to evaluation, as users do not only have to identify what the content is showing, but also when they happen, which can easily be missed when skipping through content~\cite{Kim2017Survey, Bach2014CubeReview}. Therefore, researchers have created interactive visualization systems that allow users to identify and navigate key events in time for domains such as spatial recordings~\cite{Lilija2020TemporalNavigation, Buschel2021Miria, Mahadevan2023Tesseract}, animations~\cite{Zhou2024TimeTunnel}, human motion data~\cite{Reipschlager2022Avatar, Li2023GestureExplorer, Jang2016MotionFlow, Dang2021GestureMap, Jang2014GestureAnalyzer, Hu2010MotionTrack}, biomechanics~\cite{Benard2017AnimationDiff},  \hl{and videos~\cite{Nguyen2012VideoSummagator, Li2021RouteTapestries, Matejka2014VideoLens, Dragicevic2008VideoBrowsing}.} These works have shown the importance of visualizing the whole data over time to help users identify key events to avoid unnecessary watching, and avoiding users from missing events to increase efficiency and usability. Specifically, 3D character animations are composed of rich spatial data describing numerous joints and kinematic chains. Consequently, digital content creation software such as Blender~\cite{blender} and Maya~\cite{autodeskMaya} provide various visualization tools to highlight and display this data, including joint locations and their trajectories over time. \hl{These works mainly focus on visualizing single pieces of content; this means that the visualizations presented in these works do not necessarily scale as more pieces of content are added. We investigate through a user study how to best design visualizations that scale for animation comparison, and in \system{}, we integrate established visualizations from animation and other domains for users to freely mix to support their comparison needs.}

\subsection{Comparing Spatio-Temporal Data}
More recently, researchers have investigated ways to directly compare temporal content such as 2D videos using computer vision techniques~\cite{Balakrishnan2015VideoDiff}, GenAI to extract and compare key events~\cite{Huh2025VideoDiff}, contrast and filter a large number of videos using metadata~\cite{Matejka2014VideoLens} translating videos into a shared latent space~\cite{Lin2024VideoMap}, and interfaces using information visualization comparison methods for video comparison~\cite{Baker2024ComparingVideo, Benedetto2024VideoComparison}. These works have shown how enabling skimming to quickly understand differences, and supporting multi-mode comparison to increase efficiency are vital for efficient comparison of the high amount of information in temporal data. However, these works focus on 2D content and do not need to consider 3D specific issues such as occlusion. \system{} is designed to provide skimming and multi-mode comparison while tackling the added challenges of 3D content through Spatial Lenses and filtering.

Most relevant to this work, researchers have investigated techniques for visualizing and comparing spatio-temporal data such as \hl{motion capture data~\cite{Malmstrom2016MoComp}}, robot motions~\cite{Wang2024MotionComparator, Desai2019Gepetto}, choreography creation~\cite{Han2025ChoreoCraft}, and gesture elicitation recordings~\cite{Dang2021GestureMap, Jang2016MotionFlow}. In contrast to 2D videos, 3D animations and recordings are commonly more easily integrated into the same 3D environment, removing the need for advanced computer vision methods for contrast and comparison and thus opening up new capabilities for visualization. \emph{Motion Comparator} provides visual comparison of generated robot motions to help users compare and identify the most efficient or legible robot motions for a specific task through motion paths and multiple views that abstract motion for visual inspection and alignment of two motions~\cite{Wang2024MotionComparator}. \emph{GestureMap}~\cite{Dang2021GestureMap} and \emph{MotionFlow}~\cite{Jang2016MotionFlow} visualize transitions between poses on flow maps to contrast motion differences of human gestures. These works have shown how interfaces that combine 2D and 3D visualizations of spatio-temporal data can aid user understanding, exploration, and comparison. 3D visualizations help users understand the high level of data inherent in 3D motion data and navigate to key points in the temporal dimension, while visualizations and abstractions that summarize the whole motion help users quickly identify differences. \system{} builds on these multi-mode approaches to visualize spatio-temporal differences between multiple animations through Temporal Lenses.

3D character animation consists of two key aspects in contrast to other spatio-temporal domains that add additional challenges to evaluation and comparison. First, timing is a critical aspect in animation and vital for their quality and believability~\cite{whitaker2013timing}. This aspect contrasts with other spatio-temporal domains such as robot motions which consider efficiency of motion~\cite{Wang2024MotionComparator}, and gesture elicitation, which is concerned with the pose transitions poses but not the timing of these transitions~\cite{Dang2021GestureMap, Jang2016MotionFlow}. Timing thus becomes a critical dimension that is not reflected in prior spatio-temporal comparison systems. Second, 3D animations are often used in contexts such as games or movies where camera positioning is of equal importance. This relationship between the camera, the character, and environment can significantly influence the perception of the animation and must be carefully considered~\cite{kerlow2009art}. As such, to properly evaluate and compare animations, users must also be able to consider and manipulate the angle from which the viewer will see the animation and whether the animation fits well into its intended environment. \system{} is designed to consider challenges unique to 3D character animations. 

\subsection{Generative Motion Tools}
Traditional character animation tools such as Maya~\cite{autodeskMaya} and Blender~\cite{blender} require animators to laboriously inspect, plan, and edit animations through complex interfaces, leaving little room to create variations for comparison and creative exploration. Recent advances in GenAI have enabled the generation of high-quality character motions with a wide variety of inputs such as text prompts~\cite{jiang2024motiongpt, Cong2024laserhuman}, poses and motion trajectories~\cite{Kim2022Cmib, Wan2025TLControl, Lu2022ODMO, LucasPoseGPT2022}, style or action tags~\cite{LucasPoseGPT2022, Lu2022ODMO, Jang2022MotionPuzzle}, or reference images~\cite{Jiang2025MotionChain, LucasPoseGPT2022}. These models allow animators to create a wide variety of human motions with little time and effort. However, comparison between the large amount of output from these models remains difficult as they are typically displayed in a grid within a single video or in separate videos, making it difficult to understand the temporal and spatial differences due to the lack of a common frame of reference.

The capabilities of generative motion have also recently been explored to support novel generation interfaces~\cite{Oh2025MoWa} and creative domains such as dance choreography~\cite{Liu2024DanceGen}. These interfaces are limited in that they only generate one animation at a time, or visualize animations in isolation. These interfaces do not provide a structured way for comparison and thus do not fully leverage GenAI's capability of generating varieties to aid exploration and sensemaking. In this work, we investigate techniques to evaluate generated 3D character animations to support creativity and exploration while helping animators select the best one for their needs aligned with their current workflow and practices.
\section{Formative Interviews}
With the introduction of GenAI, comparison between animations will become more relevant for animators in the future. To design tools that best suit animators, and understand current practices and limitations of comparing 3D character animations, we conducted a formative study with animators experienced in creating and reviewing 3D character animations. We were interested in how they decide whether an animation is deemed good for its purpose and how animation review is included in their workflows.

\subsection{Method}

We recruited five 3D character animators (2 female, 3 male, mean age=35, SD=12.1) via email recruitment through professional networks. We compensated the participants with 75 USD for a 1-hour remote semi-structured interview conducted via Zoom. 
The participants represented a diverse range of experience levels and production contexts. \hl{This included a hobbyist working primarily with personal film projects (P1, 3 years of experience), an early-career animator with experience in gameplay animation for games (P3, 2 years of experience), and three experienced professionals working in industry settings. These professionals included a senior animator with experience in gameplay and acting animation for games (P2, 10 years of experience), a senior animator with feature film experience (P5, 10 years of experience), and an animation director with feature film production experience (P4, 22 years of experience).}

We started the interview with participants signing a consent form and filling out a questionnaire about their demographics and 3D character animation experience. We then conducted a 1-hour  interview focusing on understanding how they compare animations in their workflows. The interviews consisted of questions about 1) the role of animation review and comparison in their workflows; 2) their criteria for what makes an animation good; and 3) what tools and methods they use to inspect and review animations\footnote{See the supplemental material for interview questions.}. 

\subsection{Findings}

Through thematic coding of the interview responses, we identified the following findings: the role of review and comparison in animators' workflows, the criteria for reviewing animations, and the methods for reviewing animations, including camera, spatial, and temporal control.

\subsubsection{Animation Review and Comparison}

Reviewing animation is a continuous process throughout animators' workflows. Due to the time and effort required for 3D character animations, the structure of an animation and its scene is usually decided early in the storyboard. As such, comparisons between animation variations are rare, but, if needed, they are assumed to be similar in action and from the same camera angle. P1 mentioned \emph{``Most of the time, I figure out what I want each of the shots to look like in the storyboarding process. So I don't have two shots look fundamentally different from each other''}. Similarly, P5 stated \emph{``Once you have your idea and you have blocked it out, that is what we stick with''}. However, animations may differ in timing or how an action is performed. In such cases, comparisons are usually made by playing animations sequentially, as stated by P4 \emph{``We do not necessarily have to play them side by side - we will just look at one''}, or by \emph{``toggling''} (P2) between versions. Because animations are typically exported as videos for reviewing, these comparisons lack spatial integration or variations within the same scene. As a consequence, comparing 3D character animations is more akin to comparing separate 2D videos than evaluating 3D animations within a shared scene.

\subsubsection{Animation Evaluation Criteria}\label{sec:criteria}

All participants emphasized that \emph{``subjective''} aspects that affect the perceived \textbf{naturalness} of the animation like \emph{``appeal''} (P4), \emph{``clarity of emotion''} (P2), \emph{``intent''} (P2), and ``believable'' (P4) character motions take precedence over identifying technical artifacts, such as foot sliding and clipping issues while judging an animation and its suitability. Animators also noted that these factors are purely assessed by visually inspecting the animation without relying on specific metrics, as stated by P1: \emph{``It is all in the gut''}. Technical principles such as \textbf{timing} and \textbf{spatial} components, but also arcs and weight were also mentioned as crucial for high-quality animations. Participants again visually examine these elements with or without additional visualizations to ensure that the animation adheres to established standards and style guides of the project. However, the quality of the underlying ideas and creative vision are often prioritized over minor technical artifacts that can be addressed later in the production process. 

\subsubsection{Camera Control Practices}
When reviewing animations, all participants mentioned that the review is conducted in context from the camera's perspective of the specific shot where the animation will be situated as stated by P5: \emph{``I review in the camera view, since for me, the camera is the last thing that people are going to see or the only thing people are going to see.''} The type of animation or the context does not significantly impact the decision to review in camera. P2 mentioned: \emph{``Even for a game where we work on multiple angles, there is no need for the animator to actually manipulate the viewport to see different angles''} \hl{highlighting how specific angles are prioritized for the animation and other angles are only considered during the final polish.} Participants would therefore make sure that before any animation is performed, they would ensure that one or multiple cameras are setup correctly. P1 said \emph{``I mostly set up my cameras first by what they look like in the storyboards''} and that inspecting animations in the viewport may be misleading due to varying \emph{``focal lengths''} or \emph{``compositions''}. This was further emphasized by P5 who stated \emph{``You need to make sure that you stick to that camera and that camera angle is good.''} All participants mentioned that setting the camera angle saves time, as only the visible parts need to be proficiently animated. P3 stated: \emph{``You have a camera angle, and then everything would be animated to that particular angle. So it could look horrible from the back.''} To ensure a good camera angle, animators stage their shot with placeholder objects to ensure that the animation is correctly positioned within the environment. As said by P3 \emph{``having good staging really helps, you know, what place you set the camera and what props are surrounded by it.''}

\subsubsection{Spatial Control Practices}

Participants (5/5) mentioned using different spatial visualizations to find animation issues and visualize arcs and movement smoothness. Most often, animators identify issues from the intended camera view and then further inspect in the viewport. Participants prefer a \emph{``plain''} camera view to inspect animations holistically before focusing on specific parts, as stated by P4: \emph{``We will always start holistically. But, if something does not feel right, we will isolate it.''} Participants would in various ways isolate specific body parts or features of the animation by, for example, attach objects to joints to accentuate the joint, turn off parts of the animation to reduce visual clutter, or inspect motion trails of joints and chains. Participants expressed that being able to include and toggle such visualizations in the scene view would be helpful for reviewing, as these visualizations are mainly limited to the viewport and not in the exported videos from the intended camera angle. Participants also mentioned that scene objects and camera placement help to review spatial aspects to ensure that the animation is scaled correctly.

\subsubsection{Temporal Control Practices}
All participants commented that temporal control is essential for inspecting and experimenting with animations timing and speeds, typically by scrubbing the timeline or playing at different speeds. P2 mentioned that \emph{``a lot of things are either floaty or too fast. [..] You can play back at different speeds so that I use that.''} This was further emphasized by P4: \emph{``Playback speed is a nice way just to experiment. Say this feels really slow. Let us just play it at 36 frames a second, just to see.''} Participants expressed the need for more control over playback speeds to make targeted adjustments to specific sections of animations and highlighted the importance of different playback modes such as looping, stepping through frames, and various playback speeds. Finally, participants emphasized that scene object references, such as a character walking past a reference object, are helpful for evaluating timing. 

\subsection{Design Goals}
Guided by our formative interviews and previous work on comparing temporal and spatial-temporal data such as videos~\cite{Huh2025VideoDiff} and robot motions~\cite{Wang2024MotionComparator}, we established four design goals to guide comparison of 3D character animations.

\begin{description}
    \item[DG1] \textbf{Contextualize the animation in its intended environment.} All animators multiple times stated the importance of seeing the animations in their intended context and that they would block out their scene and camera before creating the animation. Current GenAI models~\cite{tevet2023mdm, shafir2024human} and animation libraries such as Mixamo~\cite{mixamo} typically display animations in plain and empty scenes, making it difficult to understand and compare spatial and timing aspects. Adding animations to their intended camera view and scene through scene blocking tools would help animators make more informed choices about animation suitability and quality.    
    \item[DG2] \textbf{Provide a temporal overview of the whole animation.} The temporal aspect of animations makes it difficult to skim differences, as they are likely misaligned and you can only see one frame at a time~\cite{Huh2025VideoDiff, Wang2024MotionComparator}. This can become time-consuming and tedious as the number of animations to compare and their individual lengths increase. Techniques that visualize these differences over time can help users quickly understand the differences and make informed decisions.
    \item[DG3] \textbf{Support comparison via multiple modes.} 
    All animators mentioned how they use different tools depending on their current needs. Viewing differences in only one mode such as timeline visualizations or 3D spatial visualizations are insufficient as animators consider multiple dimensions of spatial, temporal, and aesthetics when evaluating animations. Furthermore, the type of animation and its movement may affect the suitability of tools. Providing multiple modes helps users select the best tool for their desired comparison.
    \item[DG4] \textbf{Enable filtering to visualize data of interest.} 3D character animations consist of a large amount of detailed spatial data, which can easily occlude each other and overload the user. All animators mentioned that although looking at animations holistically is important, they frequently focus on key joints and parts of the animation. Supporting filtering of unwanted data and highlighting key information is important to reduce cognitive load and aid decision-making.     
\end{description}

\section{\system{}}

\begin{figure*}[t]
    \centering
    \includegraphics[width=1\linewidth]{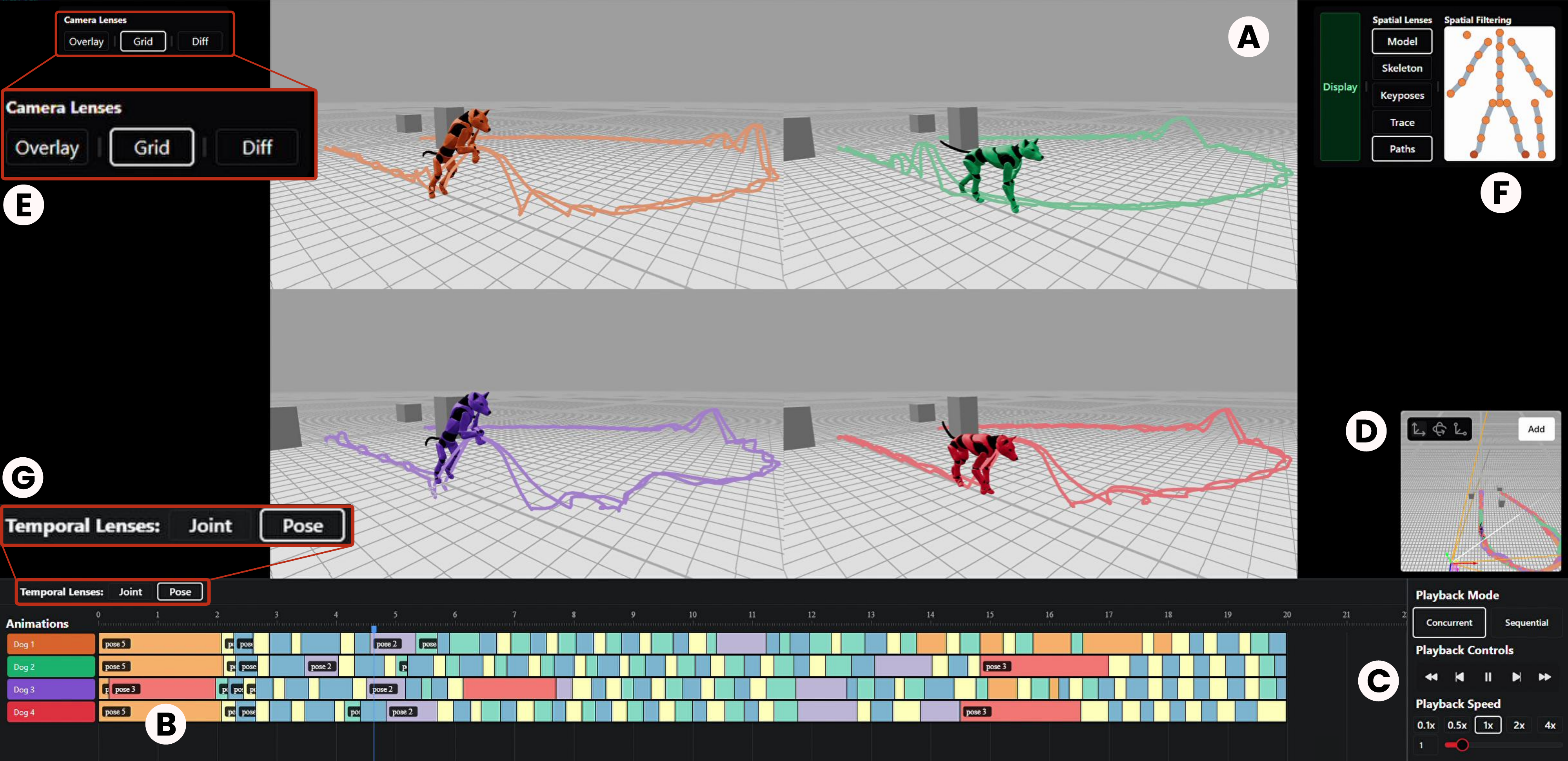}
    \caption{\system{} overview. Users can see the played animations  in the main camera view (a). Users can further see the four loaded animations in the timeline view and scrub the timeline slider (b). Playback controls control how animations are played (c). The scene controls allows users to change the main camera angle and add reference objects to the scene (d). Users can change the Camera Lens for different comparative modes in the main camera view (e). Users can enable and filter Spatial Lenses in the camera view with the Spatial Lenses menu (f). Users can change visualizations on the timeline bars through the Temporal Lenses menu for easy overview and comparison of the whole animations (g).}
    \label{fig:system-overview}
    \Description{This figure shows a screenshot overview of the interface. The different components are marked. A: In the center are 4 animations of a dog character performing a running a circuit. B: In the bottom there is a timeline of all the four animations which the users can edit. C: In the bottom right there is a menu containing playback controls and a toggle for different playback modes (concurrent, sequential). D: Above the playback controls there is a thumbnail of the scene from a different camera vantage point for scene editing. E: There is a top left sidebar that shows a toggle between three camera lenses (Overlay, Grid, Diff) each visualized. On the top right there is a menu showing a toggle of spatial lenses (model, skeleton, keyposes, trace, path), and a skeleton visualization which users can select to show or hide individual joints. G: Above the bar bottom to the left the users can toggle between different temporal bar visualizations (Pose, Joint)
}
\end{figure*}

\system{} is a web-based visual comparison tool (\autoref{fig:system-overview}) designed to support efficient comparison of alternative 3D character animations created from a generative model. \system{} supports easy comparison between alternatives by situating animations in their intended environment and camera angle, aligning animations on the timeline, quick switching between Camera Lenses for comparison, highlighting differences at a glance through Temporal Lenses, and highlighting spatial differences through Spatial Lenses. Users can filter and organize information by selectively filtering animations and their affiliated visualizations. \hl{\system{} is designed with the assumption that animations are represented as continuous sequences of poses (e.g., joint positions over time), which is common in generative and motion capture pipelines, but also generalizes to keyframe-based animations after interpolation.}

\subsection{Interface}
When uploading a set of animations, users can see all the animations in the main camera view (\autoref{fig:system-overview}a) and on the timeline (\autoref{fig:system-overview}b). Users use playback controls (\autoref{fig:system-overview}c) to control which and how animations are played. With scene controls, users adjust the camera position and angle used in the main camera view and add and manipulate 3D objects (\autoref{fig:system-overview}d) as scene references for in-context comparison. Users can compare the differences between animations by skimming the overall differences through \textbf{Temporal Lenses} on the timeline (\autoref{fig:system-overview}g, \autoref{sec:temporal-lens}), visualizing spatial movements of the animations through \textbf{Spatial Lenses} (\autoref{fig:system-overview}f, \autoref{sec:spatial-lens}), or changing the \textbf{Camera Lenses} to show animations overlaid or side-by-side (\autoref{fig:system-overview}e, \autoref{sec:camera-lens}). To control the information displayed, users can filter the animations to inspect and the spatial data to include in the camera view (\autoref{fig:system-overview}f).

\subsubsection{Scene Controls}

\begin{figure}[t]
    \centering
    \includegraphics[width=\linewidth]{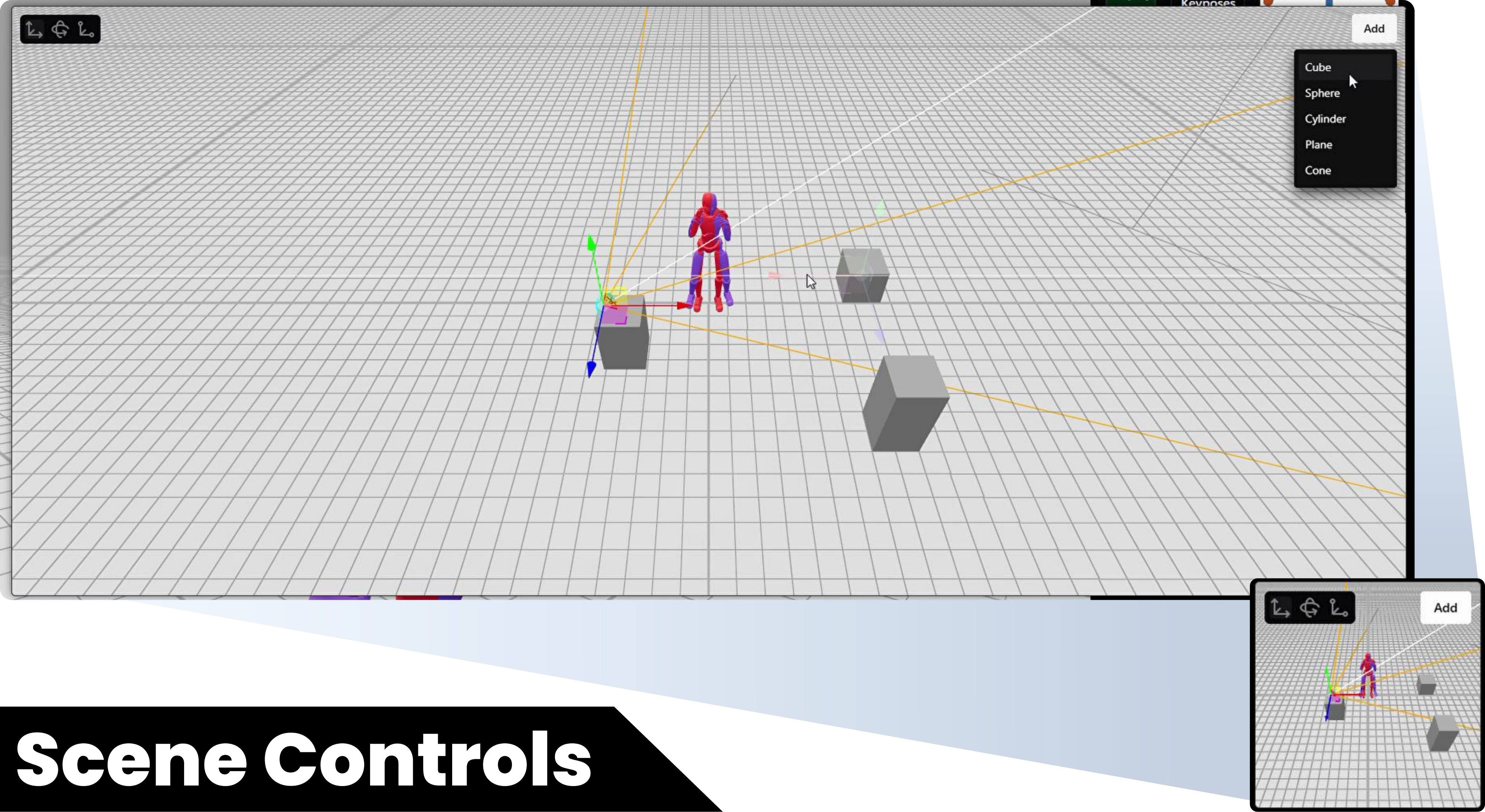}
    \caption{Users use the contextual scene controls to set up the environment in which the animation will be used in. Users can add objects to the scene, and change their position, rotation and scale. Users can also change the main camera position and rotation. }
    \label{fig:contextual-controls}
    \Description{This figure shows a zoomed in screenshot of the scene controls. Where users can move the camera, and add and manipulate scene objects to build a scene. The image consist of a scene with 4 boxing characters. There are 1 camera and 3 cubes highlighting the scene view and blocking items which the user can move and edit. There is also a drop down menu to add additional blocking items with primitives such as cubes, spheres, cylinders, planes and cones.
}
\end{figure}

Animators stated multiple times during the formative interviews that seeing the animation from the intended camera angle and environment is essential for accurate review and comparison (DG1). As such, users are expected to set the position and angle of the main camera using the scene controls (\autoref{fig:system-overview}d) before visual evaluation and comparison of the animations in the main camera view (\autoref{fig:system-overview}a). These scene controls consist of a viewport with an adjustable size for scene creation and editing (\autoref{fig:contextual-controls}). From this viewport, the user can adjust the position and rotation of the main camera to set up their shot, which adjusts the angle at which the user sees the animations in the main camera view (\autoref{fig:system-overview}a). Users are also able to add and place 3D primitives, such as cubes, spheres, and planes. This enables the user to block out their scene to aid in the review and ensure that the animations are well integrated to its intended environment and adhere to the spatial limitations of the scene.

To aid inspection, users can also change the camera angle and position in the main camera view by dragging the left mouse button for rotation, the right mouse button for panning, and the scroll wheel for zooming. When the camera angle is changed in the main camera view, a reset button appears to help the user return to the main camera angle that they have set in the scene control viewport.

\subsubsection{Playback Controls}

\system{} includes typical playback controls found in video and animation editor tools (\autoref{fig:system-overview}c). Users can play and pause, step through frame-by-frame, and jump to the start and end of animations. Users can also change the playback speed to speed up to make visual inspection faster, or slow down to ensure that they can see all details. Users can also drag animations in the timeline (\autoref{fig:system-overview}b) to offset and align animations for timing and pose-to-pose spatial comparisons. Users can select and deselect animations to include or exclude them from the scene and playback to handle information overload and focus on specific animations (DG4). Finally, \system{} consists of two playback modes to assist comparison (DG3). The \textbf{Concurrent} playback mode plays all selected animations at the same time, while the \textbf{Sequential} playback mode plays the selected animations one after each other. 

\subsection{Comparison with Camera Lenses} \label{sec:camera-lens}

In addition to simply viewing the animations from the main camera perspective, \system{} provides three \textit{Camera Lenses} (\autoref{fig:system-overview}e) to switch between superposition, juxtaposition, and explicit-encoding comparison of animations (DG3). These techniques leverage the fact that 3D animations can be integrated into the same spatial environment and support easy switching between the different comparison modes. Furthermore, users are able to select which animations to show in the Camera Lens by selecting or deselecting their respective bars on the timeline (DG4). 

\begin{figure}[t]
  \centering
  \includegraphics[width=\linewidth]{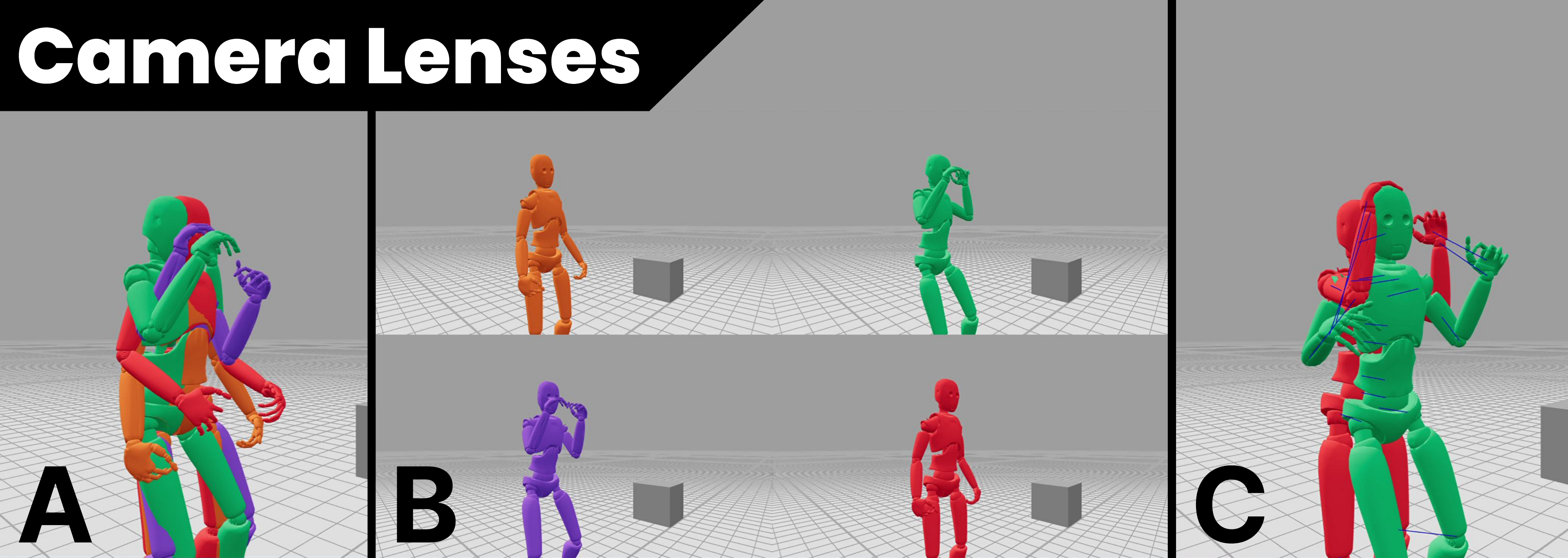}
  \caption{Camera Lenses. (a) Overlay for superposition comparison. (b) Grid for side-by-side comparison. (c) Diff to visualize joint position differences between two animations, indicated by lines connecting them.}
  \Description{Figure 4: This image show screenshots of all Camera Leneses side-by-side. A: Overlay where 4 character animations of the character is boxing are rendered on top of each others. B: Grid showing 4 animations side-by-side in a 2x2 grid of the characters performing their boxing animations. C: Diff where two animations are rendered on top of each other, and lines are drawn to each characters respective joint to accentuate differences in joint positions.
}
  \label{fig:camera-lens}
\end{figure}

The \textbf{Overlay} Camera Lens (\autoref{fig:camera-lens}a) overlays all selected animations to be rendered on top of each other for superpositioned comparison from the main camera perspective. This puts the animations into a common frame-of-reference to enables the user to easily see spatial differences between the animations. Users can select and deselect animations in the timeline to handle information overload and focus only on the animations of interest.

The \textbf{Grid} Camera Lens (\autoref{fig:camera-lens}b) puts all animations in separate views, all from the main camera perspective side-by-side in a grid for juxtaposed comparison. This enables users to view animations in isolation without animations occluding each other, while also enabling comparison. Animations can be selected and deselected in the timeline to include them or exclude them in the grid.

Finally, the \textbf{Diff} Camera Lens (\autoref{fig:camera-lens}c) enables users to contrast two animations through an explicit-encoding approach. The animation will overlay the animations into the same space as the overlay and in addition draw lines between the corresponding joints of the animations in the current frame to highlight the differences between the joint positions in space. To avoid clutter and occlusion, this Camera Lens is limited to two animations at a time, selected through the timeline. 

\subsection{Comparison with Spatial Lenses}\label{sec:spatial-lens}

\begin{figure}[t]
    \centering
    \includegraphics[width=1\linewidth]{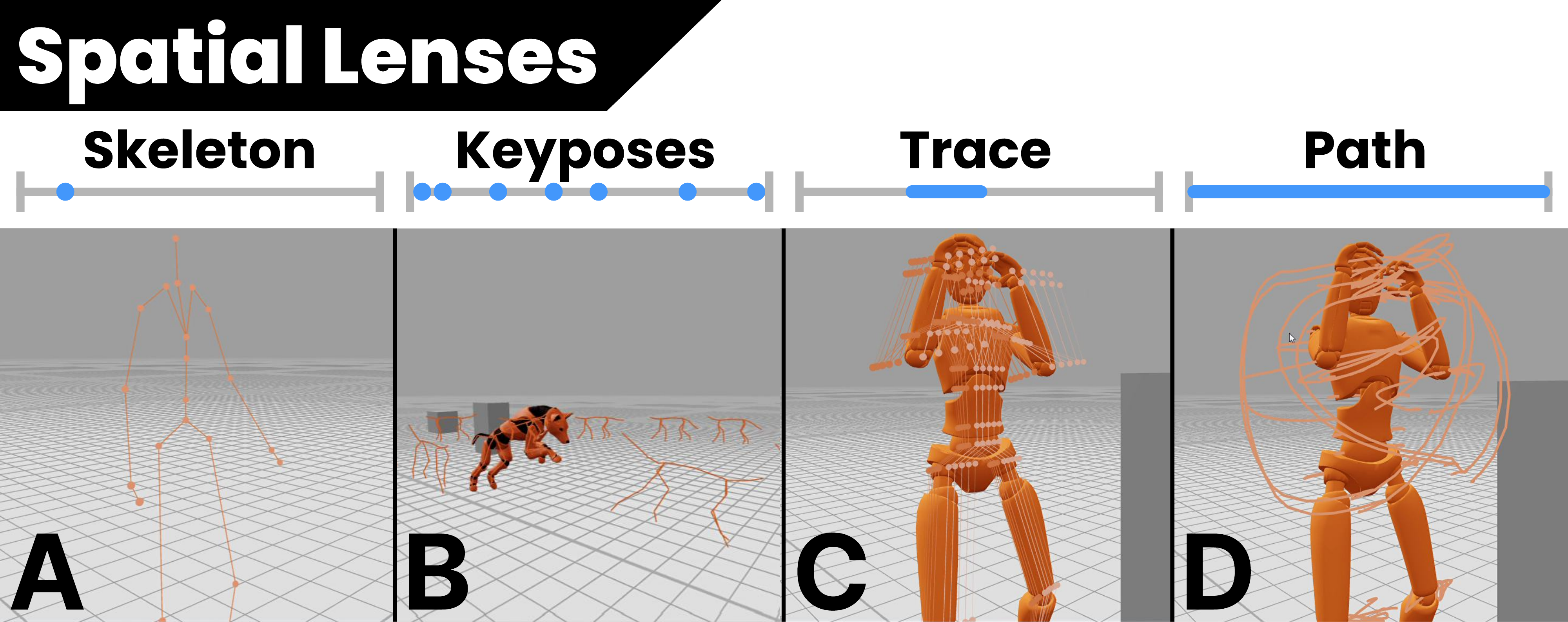}
    \caption{In addition to displaying the character model, \system{} provides four Spatial Lenses to visualize the animations spatially. Each Spatial Lens visualizes a different time interval of the animations (blue). Skeleton (a) visualizes the current frame. Keyposes visualize multiple frames (b). Trace visualizes the time interval before and after the current frame (c). Path visualizes the whole animation (d).}
    \Description{Figure 5: This image shows screenshots of all spatial lenses side-by-side. A: Skeleton showing a skeleton of the character and its joints. B: Keyposes shows the keyposes in skeleton format overlaid on top of the scene. C: Trace shows the positions and bones of the skeleton in the preceeeding and subsequent frames rendered on top of each other. D: Path shows the the joint paths of all the joints. Showcasing its whole movement across the animation. 
}
    \label{fig:spatial-lenses}
\end{figure}

\begin{figure}[t]
    \centering
    \includegraphics[width=.8\linewidth]{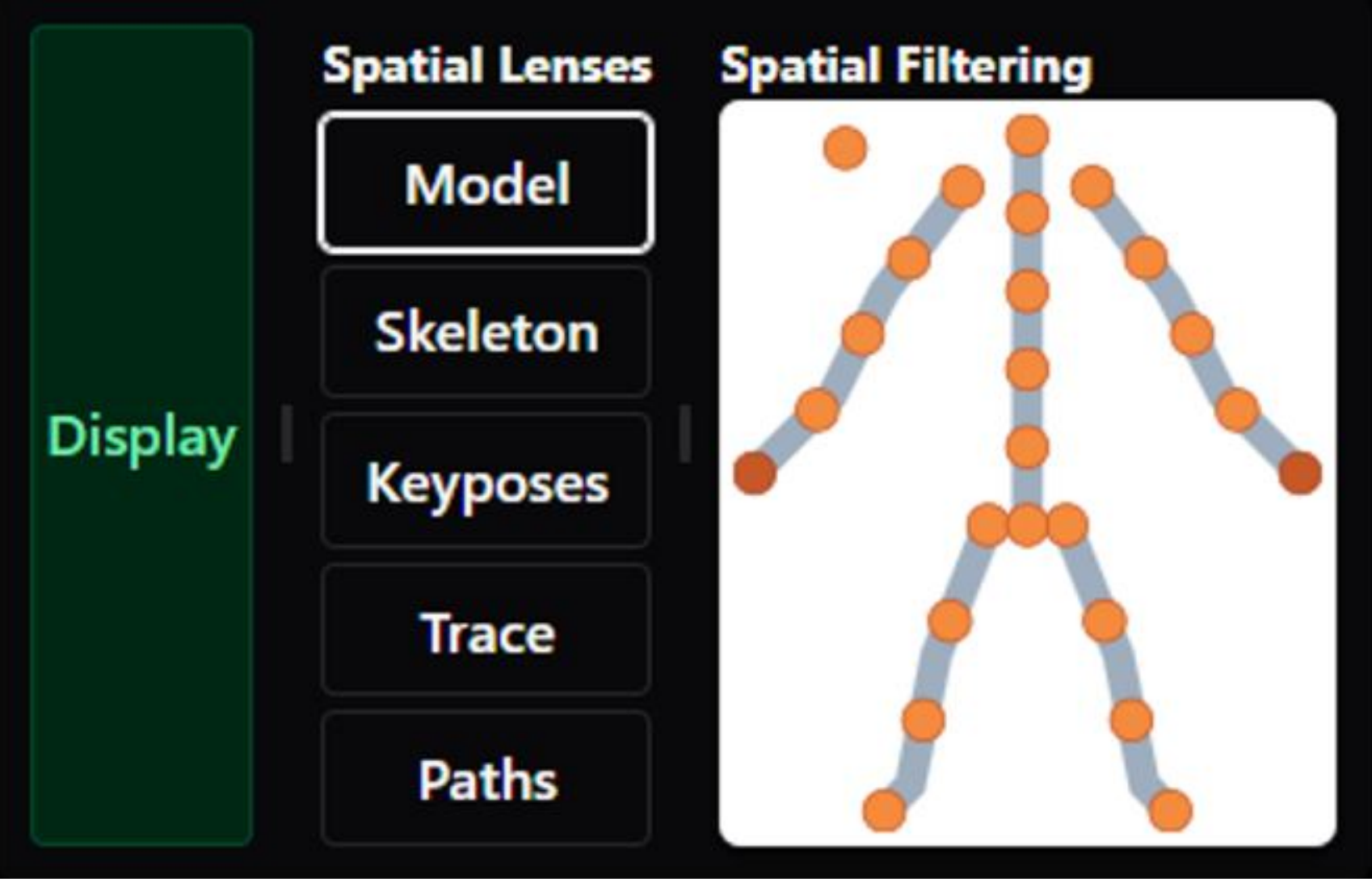}
    \caption{The spatial menu can be used to toggle the Spatial Lenses and select which joints or chains should be visualized. In this example, only the left and right hand are selected and would be visualized in the scene.}
    \Description{Screenshot of the spatial menu showing how users can select joints to visualize or hide them. To the left, There is a large green button to display or hide all visualizations. In the middle, there is a row of buttons to toggle different visualizations (Model, Skeleton, Keyposes, Trace, Paths). To the right, There is a visualizations showing the skeleton and its joints and bones which the user can click on to visualize or hide specific parts of the skeleton. 
}
    \label{fig:spatial-filtering}
\end{figure}

\system{} also employ \emph{Spatial Lenses} (\autoref{fig:system-overview}f) that visualize the spatial positions and movements of the animation skeletons integrated into the main camera view to assist users in evaluating and comparing the animations (\autoref{fig:spatial-lenses}). The Spatial Lenses are based on established visualization techniques commonly used in 3D animation software~\cite{autodeskMaya, blender} and differ in the time intervals they visualize. These visualizations are commonly used in character animation to evaluate single animations
by accentuating joint positions and decluttering the character model. However, their efficacy for comparison remain unknown. \system{} enables users to quickly toggle or combine the Spatial Lenses with each other or other components to suit current comparison needs (DG3) and select which parts of the character to visualize to highlight key information or to filter unwanted parts of the skeleton (DG4).

\hl{The \textbf{Skeleton} Lens (\autoref{fig:spatial-lenses}a) visualizes the character’s joint positions and the bones connecting them in the current frame.} By turning off the character model and displaying the skeleton, users can also visualize the animations in a less occluded manner. 

The \textbf{Keyposes} Lens (\autoref{fig:spatial-lenses}b) visualizes the keyposes extracted from the animation and places them in the scene. This gives users an overview of the animation at key frames during the animation at the same time. For data that does not generally include  keyposes, such as motion capture data or GenAI-created animations, we use the algorithm proposed by \citet{Tanuwijaya2010Keypose} to extract $k$ number of keyposes from the animation. 

The \textbf{Trace} Lens (\autoref{fig:spatial-lenses}c) visualizes the $n$ past and future frames of the joint and skeleton positions, like the traditional onion skin technique. This enables users to evaluate the smoothness of the animation and see a larger time interval, helping users align animations and reducing the need for precise  alignment of animations for comparisons.

The \textbf{Path} Lens (\autoref{fig:spatial-lenses}d) visualizes the full path of the joint in space throughout the entire animation (DG2). This enables the user to see an overview of the full range of motion of the joint at any time, without having to scrub through the animation. As such, spatial comparisons between animations can be made without having to align the animations. Furthermore, this also enables the user to see whether a joint at any time collides with the environment during the environment without any playback, to see how well the animation is integrated to its intended context (DG1).

Due to the high number of joints and the detailed 3D data encoded into each joint, being able to filter and select information to visualize only the key information of interest is essential (DG4). To do so, \system{} provides a menu where users can toggle spatial visualizations, and a joint map (\autoref{fig:spatial-filtering}) where users can select and deselect the specific joints the user wants to visualize. Users are also able to select chains (e.g., left arm) to visualize or hide all joints that are part of that chain. This capability enables users to highlight specific parts of the animation in which they are interested while hiding other parts to minimize occlusion. 

\subsection{Comparison with Temporal Lenses}\label{sec:temporal-lens}

To give users an overview of all animations (DG2) the system contains \emph{Temporal Lenses} which visualize how the animations change over time on the timeline bars (\autoref{fig:system-overview}b). These enable users to understand how the animations progress over time, contrast differences between animations without needing visual inspection and \hl{reduce the need for manually dragging animations on the timeline for temporal alignment}. Users can select between two different visualizations (\autoref{fig:system-overview}g).

\begin{figure}[t]
    \centering
    \includegraphics[width=1\linewidth]{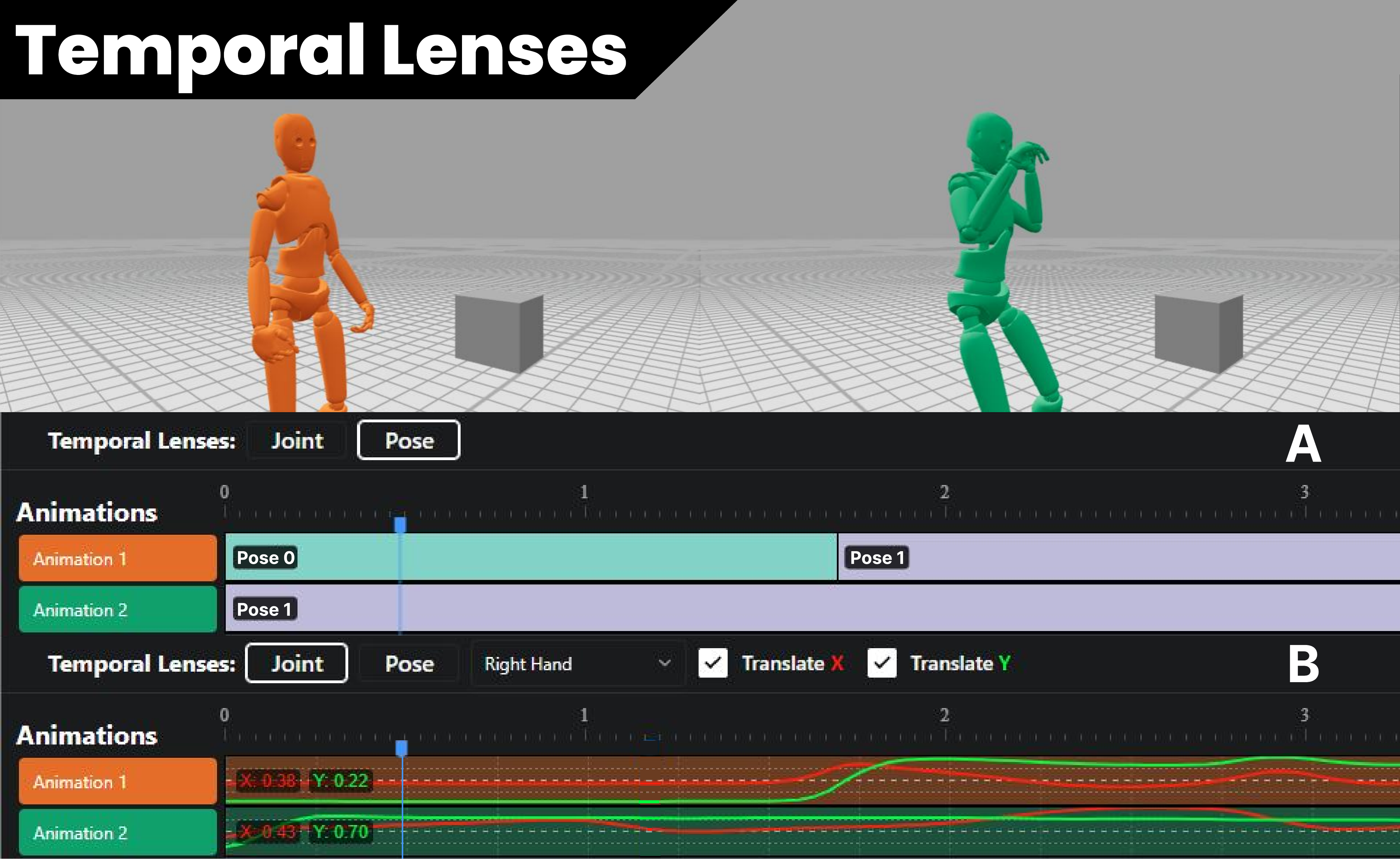}
    \caption{The Temporal Lenses are visualized on the timeline bars of each animation and give an overview of the whole animations. The Pose Lens (a) visualizes the transition of poses during the animation. In this case teal represents an idle pose, while purple represents an arms up boxing pose. The Joint Lens (b) visualizes the x and y position in the main camera space of a specific joint over time. In this case it allows the users to see when the right hand is selected and allows the user to see when it raised or lowered.}
    \Description{This image show a screenshot of all Temporal Lenses side-by-side. At the top there are two character performing a boxing animation side-by-side. Then the pose temporal lens is shown it has clustered poses into two bars along the timeline. One animation starts with an idle pose, while the other animation starts with a boxing pose. They therefore have two different bar colors. When the first animation transitions into a boxing pose, they have the same color on the timeline. At the bottom shows the Joint temporal lens, the user has selected the right hand joint, and along the timeline it shows the x any y position of each animations joint. The user can see and compare the joint movement between then animations. 
}
    \label{fig:temporal-lens}
\end{figure}

The \textbf{Pose} Lens provides an overview of the distinct poses performed across all animations by clustering similar poses together and then displaying the clustered poses as segments on the timeline bars of each animation (\autoref{fig:temporal-lens}a). This is achieved by clustering to identify similarities between animation poses. When the slider is moved to the start of the animation, users can observe clusters such as \textit{“Pose 0”} representing an idle position, and \textit{“Pose 1”} representing an arms-raised posture. The clustering is inspired by works on the comparison of gesture ellication data~\cite{Jang2016MotionFlow, Dang2021GestureMap}. \hl{An average animation is computed using Dynamic Time Warping Barycenter Averaging (DBA) to align and aggregate temporal sequences. Each animation is represented as a sequence of joint positions. DBA initializes with a reference sequence (randomly selected among candidates), aligns all animations to it using Dynamic Time Warping (DTW) to account for temporal variation, and updates each frame by averaging the aligned poses across sequences. This process iterates until convergence, yielding a temporally normalized mean animation that preserves the core motion while reducing timing differences.}

This average animation is then used for X-means clustering, grouping its poses into distinct groups. \hl{Clustering is performed on joint positions expressed relative to the root joint, making the representation invariant to global position and orientation. At the same time, because relative joint configurations are preserved, the clustering retains semantic distinctions such as which limb performs an action (e.g., left vs. right arm). The x-means cluster will end up with $N$ distinct pose groups. We then perform a second k-means clustering using the same number of clusters ($N$) as decided by the initial x-means clustering and its derived clusters as initial centroids. As input we use all poses from the animations using the same joint representation as in the original clustering.} During this second clustering, all animation poses are assigned to their respective clusters. These clusters are then visualized as color-coded segments on the animation time bars. Visualizing the clusters as segments on the animation time bars enables users to see and compare the exact timings of pose transitions and how the animation progress over time.

The \textbf{Joint} Lens enables visualization of joint positions within the main camera view's 2D coordinate system throughout the animation (\autoref{fig:temporal-lens}b). When the Joint Lens is activated, the timeline bars of each animation display curves of a joint's X and Y movements in camera space throughout the entire animation. Users select the joint (e.g., Right Hand) to visualize using a dropdown menu and toggle to display X and Y coordinates. Joint positions are projected into camera space, normalized using the global minimum and maximum values of all animations to maintain a consistent scale, and then clamped to the camera boundaries to indicate when joints enter or exit the camera view. This functionality allow users to inspect when and how the joints move throughout the animation for a more detailed evaluation and comparison. If the position of a joint extends beyond the bounds of the camera, this is indicated by the position curves moving beyond the bounds of the timeline bar of the animation. \hl{This enables the user to contextualize the animations in the intended camera shot (DG1) for quick overview how the animations move within the shot. For future versions, this the Joint Lens can be integrated with object-space graph editors as is common in animations tools such as Maya for animation editing. This would allow animators to make edits through their established tools and see how that affects in the animation within the shot.}

\subsection{Implementation}
We implemented \system{} as a web-based interface using React.js and Three.js. Animations could be uploaded in batches via FBX format after being generated via a generative model. Animation clustering and processing were handled by a backend Flask-based python server. All animations were played in 24 frames per second. The following parameters for techniques and visualizations were $n=10$ for Trace, and $k=15$ for Keyposes. The current implementation is limited to only static camera shots.
\section{User Evaluation}
To understand how \system{} assists users in comparing character animations and uncover what design guidelines and tool designs are usable for animation comparison, we conducted a study with 12 participants experienced in 3D character animation. The participants performed a set of comparative tasks in four scenarios with different types of character animation. Our study aimed to evaluate the perceived usability and utility of \system{} in supporting different animation comparison tasks, identify the strategies that users adopt, identify the most useful visualization and tool designs, and understand how different animation types may affect their comparison strategies. As the formative study showed animation comparison to be rare and mainly through playing animation videos one-by-one, we opted to not include a study baseline and instead included more animation scenarios to understand what visualizations and designs are useful across various comparison and animation types to inform future design.

\subsection{Method}
\begin{table*}[t]
    \centering
    \resizebox{1\linewidth}{!}{
    \begin{tabular}{llll}
        \toprule
      Category   &  Type & Template Task Question & Example Question\\
      \midrule
       Action \& Style & Multi-select & T1. Which animation(s) perform the action \emph{\{action\}}? & \hl{Which animation(s) performed a wave with one arm?}\\
       Spatial & Single-select & T2: Which animation performs the \emph{\{action\}} in the most \emph{\{description\}}? & \hl{Which animation reached their hands the highest within the shot?}\\
        Temporal Timing & Single-select & T3: Which animation performs the \emph{\{action\}} in the most \emph{\{timing\}}? & \hl{Which animation started their waving action last?}\\
       Temporal Speed & Single-select & T4: Which animation completes the \emph{\{action\}} in the most \emph{\{duration\}}? & \hl{Which animation had the longest waving action from start to finish?}\\
       Naturalness & Single-select & T5: Which animation was the smoothest/most natural? & \hl{N/A}\\
        \bottomrule
    \end{tabular}
    }
    \caption{Task questions used in the user study included both single-select (choosing one answer that meets the criteria) and multi-select (selecting all answers that meet the criteria) formats.}
    \label{tab:study_questions}
\end{table*}

\subsubsection{Participants}
We recruited 12 participants (3 female, 9 male, mean age=35, SD=8.6) with diverse experience in creating character animations, using character animations in games or videos, or creating character animation tools. Participants were recruited through a communication platform and word of mouth by posting announcements from within the author institutions. Two participants self-identified as experts in character animation (M= 17.0, SD=13.0 years of experience); four self-identified as proficient (M=9.0, SD=5.9 years of experience); and six identified as amateurs (M=2.5, SD=3.5 years of experience). Participants were compensated 100 USD for the 1.5 hour study. The study was approved by the internal institutional ethics review board.

\subsubsection{Materials}
\begin{figure}[t]
    \centering
    \includegraphics[width=1\linewidth]{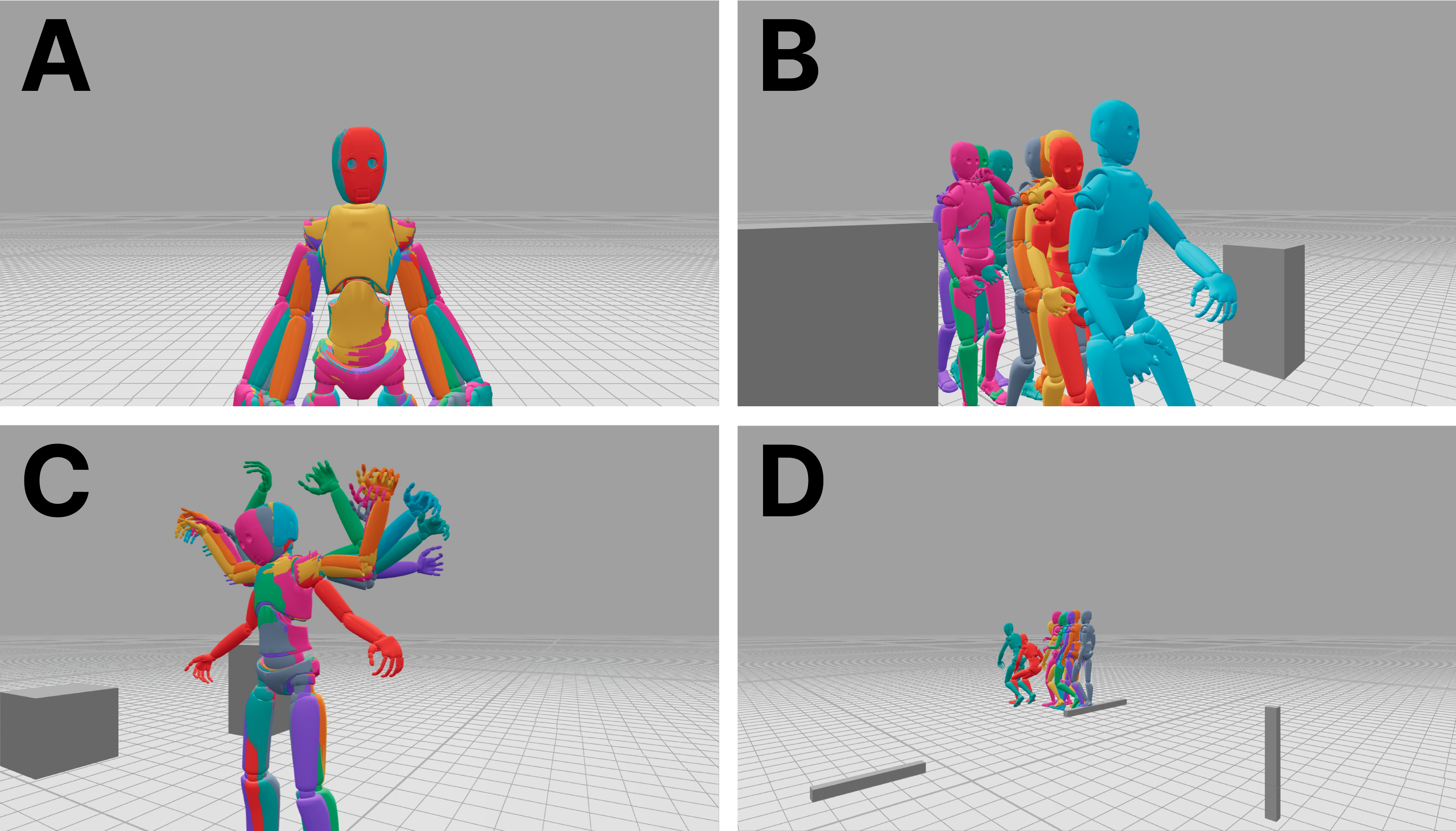}
    \caption{The four animation scenarios. (a) In the Wave scenario (a), the characters perform a short waving action. In the Watch scenario (b), the characters walk along the camera view and raise their arm to look at their wristwatch. In the Stretch scenario (c), the characters perform a stretch, followed by jumping jacks and finish with another stretch. In the Walk scenario (d), the characters walk across a course and jump over obstacles.}
    \Description{This image shows a grid of screenshots of the four animation scenarios. A: 9 animations rendered on top of each other facing the camera, and about to raise their arms to wave at the camera. B: 9 animations walking diagonally towards the camera at different pace, some animations have started to raise their arm to look at their wristwatch. C: 9 animations standing in front of the camera doing jumping jacks. The animations are at different points of the jumping jack motion with their arm. D: A far shot of 9 overlapping animations running around a course, some animations are mid-jump over an obstacle.
}
    \label{fig:study-conditions}
\end{figure}

The early pilots of this study showed that the duration of the animation and the presence of translational root motion across the screen are important factors that influence comparison. Longer animations contain more complex movements, whereas translational movement accentuates challenges created by timing misalignment. Animations without translational movement face more occlusion issues. To investigate the effect of animation types on comparison, we created four scenarios, each containing nine animations generated from GenAI models. We decided on nine animations to balance individual inspection and comparison between and beyond two animations within a reasonable time frame for our study. Each scenario varied by animation length ranging from shorter animations (6s) to longer animations (20s), and with or without the presence of translational movement. For all scenarios we used the Robot character included in the Autodesk Maya MotionMaker tool~\cite{autodeskMaya}. Each scenario had a preset camera angle and reference scene objects to maintain a consistent viewing perspective and scene for participants (\autoref{fig:study-conditions}). The participants performed the comparative tasks within each scenario.

\begin{description}
    \item[Wave Scenario:] The animations in this scenario performed a waving action (\autoref{fig:study-conditions}a) and were 6 seconds long and without root movement. Nine animations were generated via the prompt \emph{``The person is waving''} with different seeds using Human Motion Diffusion Model (MDM)~\cite{tevet2023mdm}.
    \item[Watch Scenario:] This scenario consisted of short 6 second and with translational movement animations that walked forward along the camera and looking down at a wristwatch. Nine animations were generated via the prompt \emph{``The person walked forward and stopped to look at their wristwatch''} with different seeds using MDM~\cite{tevet2023mdm}. 
    \item[Stretch Scenario:] This scenario contains animations performing stretching and jumping jacks that were 20 seconds long and without root movement. Nine animations were generated with a sequence of three prompts \emph{``The person is stretching}; \emph{``The person is performing jumping jacks''}; \emph{``The person is stretching their upper body''} with different seeds using the DoubleTake algorithm of the PriorMDM model to create long animations~\cite{shafir2024human}.
    \item[Walk Scenario:] This scenario contains animations that walk and jump around a course that are 20 seconds long and with translational movement across the camera view. The nine animations used for the scenario were generated using the Maya MotionMaker tool where the animator draw a path with breakpoints for actions such as jumping which the character runs along.
\end{description}

\subsubsection{Task and Procedure}

We started the study with participants signing a consent form and filling out a demographic questionnaire about their animation background. Next, we gave a 10-minute tutorial on \system{} with an example scenario. Participants then performed a set of comparative tasks on the first animation scenario. 
For comparison tasks, participants were asked to review, compare, and answer five questions (T1-T5, \autoref{tab:study_questions}) about the animations using \system{}. These tasks were derived from the formative study (\autoref{sec:criteria}), to focus on aspects animators consider when evaluating the quality and suitability of animations in a scene, such as naturalness, timing, and style. Tasks were designed to be diverse, including single- and multi-select questions. Furthermore, as in a similar study~\cite{Huh2025VideoDiff}, we included different comparison modes in which answers can be found looking at animations in isolation (e.g., T1) and where answers are hidden, requiring a more exploratory search and comparison (e.g., T3). For each question, we measured the completion time, answer accuracy, and interaction logs to understand participant strategies and which system components participants utilized to perform each comparison. We rated accuracy as follows: 1 for completely correct answer, 0.5 for getting half or more of the answers right for multi-select questions, and 0 for less than half were correct. \hl{Note that accuracy was not collected for naturalness (T5) due to its open-ended nature.} After completing all tasks for a scenario, we conducted a post-task survey that included ratings for perceived mental demand, performance, effort, frustration, and usefulness of the system and its components in understanding animations' differences. All ratings were on a 7-point Likert scale. Participants then repeated the tasks with the next scenario. The scenarios were counterbalanced using a balanced Latin square. After finishing all scenarios, participants answered a post-study questionnaire and took part in a semi-structured interview about their comparison strategies and the perceived usefulness of \system{}. The study lasted approximately 90 minutes.
\section{Results}
We analyzed user behavior based on task performance, interaction logs, and questionnaire responses. We transcribed the semi-structured interviews and any comments made during the study. One researcher conducted a framework analysis, organizing the data into categories aligned with the system’s workflow and features to identify patterns in usability and utility. We found the following themes from the analysis of the results.

\subsection{\system{} Effectively Supports Animation Comparison}

\begin{table}[t]
    \centering
    \resizebox{1\linewidth}{!}{
    \begin{tabular}{c|c|ccccc}
    \toprule
    \textbf{Scenario} & \textbf{Metric} & \thead{T1: Action\\Matching} & \thead{T2:\\Spatial} & \thead{T3: Temporal\\Timing} & \thead{T4: Temporal\\Speed} & \thead{T5:\\Naturalness} \\
    \midrule
      Wave & Time (s) & 117 (48) & 47 (23) & 86 (43)&  75 (47) & 129 (29)\\
       & Accuracy & .97 (.10)& .92 (.29)& .83 (.39)& .92 (.29) & N/A\\
    \midrule
      Watch & Time (s) & 174 (79)& 78 (44)& 99 (68)& 52 (25) & 122 (41)\\
       & Accuracy & .83 (.20) & .92 (.29)& .83 (.39) & .92 (.29) & N/A\\
    \midrule
      Stretch & Time (s) & 211 (68) & 83 (50)& 117 (40)& 243 (109) & 219 (80)\\
       & Accuracy & .83 (.25) & .92 (.29)& 1.00 (.00)& .75 (.43) & N/A\\
       \midrule
      Walk & Time (s) & 218 (66)& 78 (40) & 78 (34) & 44 (19) & 171 (43)\\
       & Accuracy & 1.00 (.00) & .83 (.39) & .83 (.39)& 1.00 (.00) & N/A\\
   \bottomrule
    \end{tabular}
    }
    \caption{Average task completion time and accuracy (SD) for task questions. Note that accuracy was not collected for T5 (naturalness) due to its open-ended nature.}
    \label{tab:performance}
\end{table}

\begin{figure*}[t]
    \centering
    \includegraphics[width=.495\linewidth]{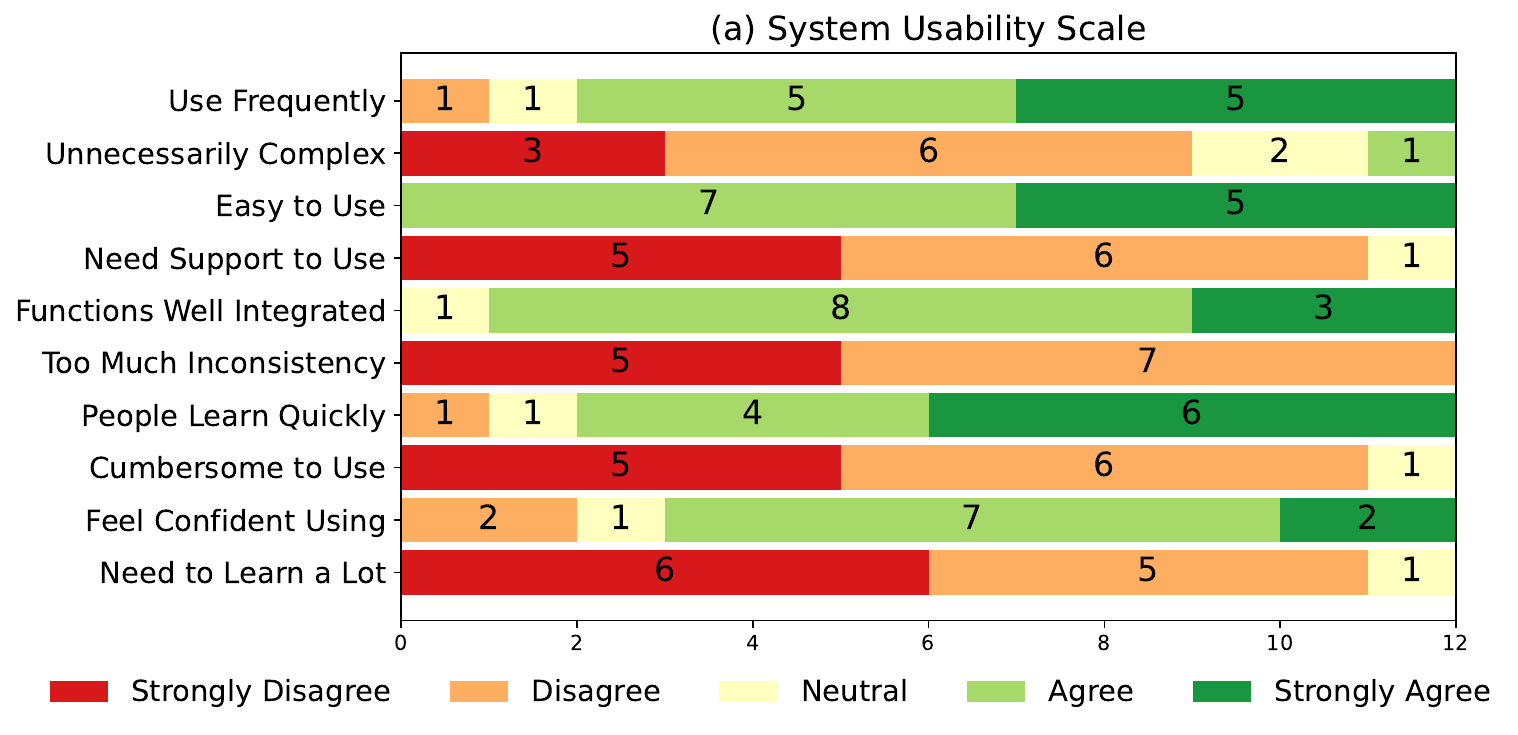}
    \includegraphics[width=.495\linewidth]{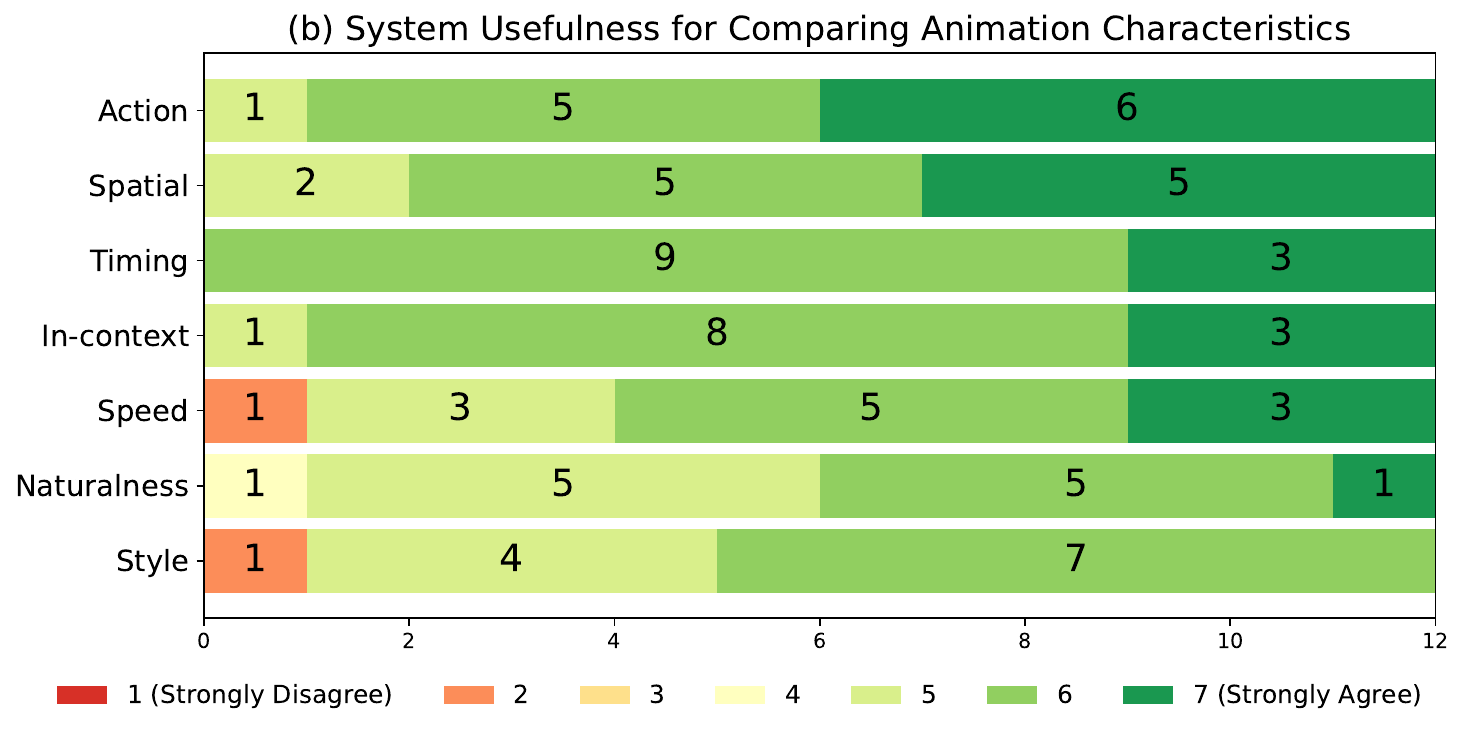}
    \caption{Subjective feedback on whole system. (a) System Usability Scale questions. (b) Participant ratings on \system{}'s perceived usefulness for different types of comparison between animations.}
    \label{fig:usability}
    \Description{Figure 9: This image show a stacked bar chart of System Usability ratings, and a bar chart of the usability ratings.
    The system usability ratings have the following results:
    Use frequently: 1 Disagree, 1 Neutral, 5 Agree, 5 Strongly Agree
    Unnecessarily Complex: 3 Strongly Disagree, 6 Disagree, 2 Neutral, 1 Agree
    Easy to Use: 7 Agree, 5 Strongly Agree
    Functions well integrated: 1 Netural, 8 Agree, 3 Strongly Disagree
    Too much inconsistency: 5 Strongly Disagree, 7 Disagree
    People Learn Quickly: 1 Disagree, 1 Neutral, 4 Agree, 6 Strongly Agree
    Cumbersom to use: 5 Strongly Disagree, 6 Disagree, 1 Neutral
    Feel Confident Using: 2 Disagree, 1 Neutral, 7 Agree, 2 Strongly Agree
    Need to learn a lot: 6 Strongly Disagree, 5 Disagree, 1 Neutral
    
    The system usefulness for comparing animation characteristics questions are ranked 1 to 7. 1 = Strongly Disagree, 7 = Strongly Agree
    Action: 1 answered 5, 5 answered 6, 6 answered 7
    Spatial: 2 answered 5, 5 answered 6, 5 answered 7
    Timing: 9 answeed 6, 3 answered 7
    In-context: 1 answered 5, 8 answered 6, 3 answered 7
    Speed: 1 answered 2, 3 answered 5, 5 answered, 6, 3 answered 7
    Naturalness: 1 answered 4, 5 answered 5, 5 answered 6, 1 answered 7
    Style: 1 answered 2, 4 answered 5, 7 answered 6
    }
\end{figure*}
All participants were able to answer the comparative questions in all scenarios with a high overall accuracy across all scenarios (\autoref{tab:performance}). In general, T1 (Action Matching) and T5 (Naturalness) were the most time-consuming questions to answer  as they required visual inspection of the animations in isolation. Meanwhile, spatial (T2) and temporal (T3 and T4) comparisons were faster, as users could rely on the Temporal and Spatial Lenses to understand differences without visual inspection. \system{} achieved a rating of $80.42 \pm 11.67$ on the System Usability Scale (\autoref{fig:usability}a), a \textit{``good''} adjective rating on the SUS benchmark~\cite{Bangor2008SUS}, with all users agreeing that it was easy to use, useful for character animation comparison, and well-aligned with their animation workflows. Participants expressed enthusiasm for integrating \system{} functionality into established digital content creation tools, stating: \textit{“I would love to see this in Maya”} (P3). In general, the participants agreed that the system could be an effective addition to their toolset describing the system as \textit{“super helpful”} (P1) and \textit{“really useful for comparing animations”} (P4). Participants developed structured strategies to manage the comparison using the flexibility of \system{} and its multiple modes. The ability to easily combine and switch between comparative modes and visualizations through the different lenses was highlighted as particularly valuable, addressing a major gap in existing tools that often require manual and time-consuming processes. As P7 explained, \textit{“My favorite feature is being able to see all of them in either Overlay or Grid at the same time. I think having that capability is really powerful”}. This was further emphasized by P12:  \textit{“Being able to visualize everything is pretty useful”}.

\subsubsection{Tools with temporal overview make efficient motion comparison by reducing  playback time and alignment.} \label{sec:time-interval}
Participants emphasized that visualizations that span the entire animation time interval, such as Pose Lens or Path, were significantly more effective for comparing animations than views limited to time points or smaller intervals. Unlike time point-based representations, which \textit{“do not capture everything”} (P2), the Path and Temporal Lenses offered a comprehensive overview of motion and captured intermediate states, reducing the need for manual alignment or frame-by-frame inspection. As one participant noted, \textit{“looking at the Path, it is much easier to see which one is higher because the keyframe does not have all the frames”} (P2). These holistic views allowed users to infer differences at a glance, such as identifying which animation is performing an action \textit{“the longest”} (P7) without relying solely on visual inspection of playback. By showing the complete trajectory, these representations also minimized the cognitive load and bias introduced by visual inspection: \textit{“Path was cool because then you can just see who moved more left or right. You don’t have to guess looking at [the animations]”} (P5). Although Skeleton, Keypose, and Trace were considered helpful in understanding single animations, participants found them less useful for comparison: \textit{“I would be interested in using those for single animations, but when comparing different animations, I just wonder what information I can get from those”} (P12). Manual alignment was also highlighted when participants mentioned the Camera Lenses, as Diff was overall not considered useful since it required significant alignment and was restricted to comparison of only two animations at a time. In general, participants valued visualizations that revealed the entire temporal context, as these supported faster and more reliable comparisons.

\begin{figure*}
    \centering
    \includegraphics[width=1\linewidth]{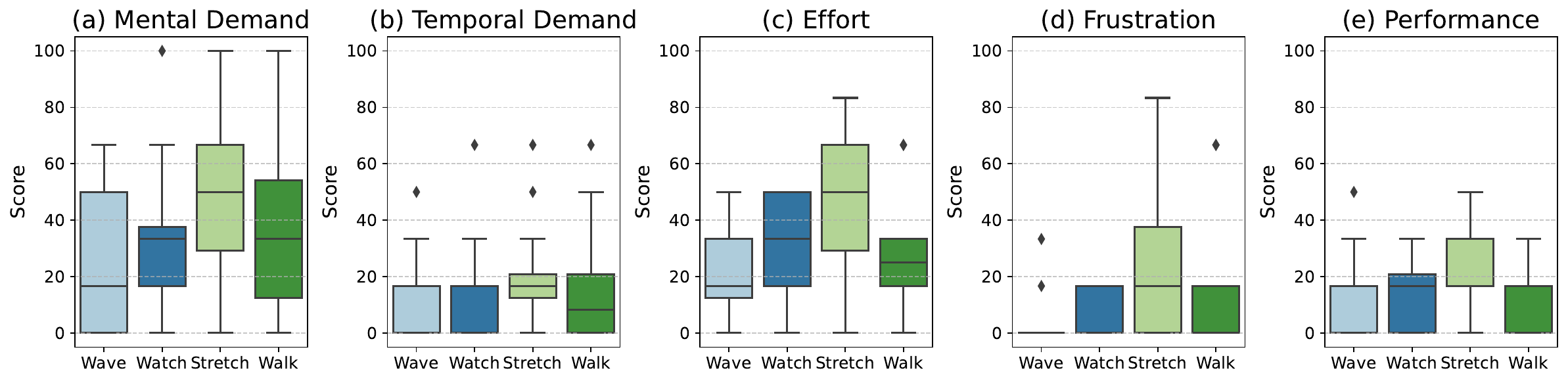}
    \caption{Raw NASA TLX answers from participants across animation scenarios.}
    \Description{
    (a) MentalDemand:
  Stretch:
    Mean: 50.00
    Box limits - Q1: 29.17, Median: 50.00, Q3: 66.67
    Whisker limits - Lower: -27.08 (actual: 0.00), Upper: 122.92 (actual: 100.00)
    Data range - Min: 0.00, Max: 100.00
  Walk:
    Mean: 38.89
    Box limits - Q1: 12.50, Median: 33.33, Q3: 54.17
    Whisker limits - Lower: -50.00 (actual: 0.00), Upper: 116.67 (actual: 100.00)
    Data range - Min: 0.00, Max: 100.00
  Watch:
    Mean: 33.33
    Box limits - Q1: 16.67, Median: 33.33, Q3: 37.50
    Whisker limits - Lower: -14.58 (actual: 0.00), Upper: 68.75 (actual: 68.75)
    Data range - Min: 0.00, Max: 100.00
  Wave:
    Mean: 26.39
    Box limits - Q1: 0.00, Median: 16.67, Q3: 50.00
    Whisker limits - Lower: -75.00 (actual: 0.00), Upper: 125.00 (actual: 66.67)
    Data range - Min: 0.00, Max: 66.67

(b) TemporalDemand:
  Stretch:
    Mean: 20.83
    Box limits - Q1: 12.50, Median: 16.67, Q3: 20.83
    Whisker limits - Lower: -0.00 (actual: 0.00), Upper: 33.33 (actual: 33.33)
    Data range - Min: 0.00, Max: 66.67
  Walk:
    Mean: 16.67
    Box limits - Q1: 0.00, Median: 8.33, Q3: 20.83
    Whisker limits - Lower: -31.25 (actual: 0.00), Upper: 52.08 (actual: 52.08)
    Data range - Min: 0.00, Max: 66.67
  Watch:
    Mean: 15.28
    Box limits - Q1: 0.00, Median: 16.67, Q3: 16.67
    Whisker limits - Lower: -25.00 (actual: 0.00), Upper: 41.67 (actual: 41.67)
    Data range - Min: 0.00, Max: 66.67
  Wave:
    Mean: 13.89
    Box limits - Q1: 0.00, Median: 16.67, Q3: 16.67
    Whisker limits - Lower: -25.00 (actual: 0.00), Upper: 41.67 (actual: 41.67)
    Data range - Min: 0.00, Max: 50.00

(c) Effort:
  Stretch:
    Mean: 45.83
    Box limits - Q1: 29.17, Median: 50.00, Q3: 66.67
    Whisker limits - Lower: -27.08 (actual: 0.00), Upper: 122.92 (actual: 83.33)
    Data range - Min: 0.00, Max: 83.33
  Walk:
    Mean: 29.17
    Box limits - Q1: 16.67, Median: 25.00, Q3: 33.33
    Whisker limits - Lower: -8.33 (actual: 0.00), Upper: 58.33 (actual: 58.33)
    Data range - Min: 0.00, Max: 66.67
  Watch:
    Mean: 29.17
    Box limits - Q1: 16.67, Median: 33.33, Q3: 50.00
    Whisker limits - Lower: -33.33 (actual: 0.00), Upper: 100.00 (actual: 50.00)
    Data range - Min: 0.00, Max: 50.00
  Wave:
    Mean: 20.83
    Box limits - Q1: 12.50, Median: 16.67, Q3: 33.33
    Whisker limits - Lower: -18.75 (actual: 0.00), Upper: 64.58 (actual: 50.00)
    Data range - Min: 0.00, Max: 50.00

(d) Frustration:
  Stretch:
    Mean: 23.61
    Box limits - Q1: 0.00, Median: 16.67, Q3: 37.50
    Whisker limits - Lower: -56.25 (actual: 0.00), Upper: 93.75 (actual: 83.33)
    Data range - Min: 0.00, Max: 83.33
  Walk:
    Mean: 13.89
    Box limits - Q1: 0.00, Median: 0.00, Q3: 16.67
    Whisker limits - Lower: -25.00 (actual: 0.00), Upper: 41.67 (actual: 41.67)
    Data range - Min: 0.00, Max: 66.67
  Watch:
    Mean: 9.72
    Box limits - Q1: 0.00, Median: 16.67, Q3: 16.67
    Whisker limits - Lower: -25.00 (actual: 0.00), Upper: 41.67 (actual: 16.67)
    Data range - Min: 0.00, Max: 16.67
  Wave:
    Mean: 4.17
    Box limits - Q1: 0.00, Median: 0.00, Q3: 0.00
    Whisker limits - Lower: 0.00 (actual: 0.00), Upper: 0.00 (actual: 0.00)
    Data range - Min: 0.00, Max: 33.33

(e) Performance:
  Stretch:
    Mean: 22.22
    Box limits - Q1: 16.67, Median: 16.67, Q3: 33.33
    Whisker limits - Lower: -8.33 (actual: 0.00), Upper: 58.33 (actual: 50.00)
    Data range - Min: 0.00, Max: 50.00
  Walk:
    Mean: 12.50
    Box limits - Q1: 0.00, Median: 16.67, Q3: 16.67
    Whisker limits - Lower: -25.00 (actual: 0.00), Upper: 41.67 (actual: 33.33)
    Data range - Min: 0.00, Max: 33.33
  Watch:
    Mean: 13.89
    Box limits - Q1: 0.00, Median: 16.67, Q3: 20.83
    Whisker limits - Lower: -31.25 (actual: 0.00), Upper: 52.08 (actual: 33.33)
    Data range - Min: 0.00, Max: 33.33
  Wave:
    Mean: 15.28
    Box limits - Q1: 0.00, Median: 16.67, Q3: 16.67
    Whisker limits - Lower: -25.00 (actual: 0.00), Upper: 41.67 (actual: 41.67)
    Data range - Min: 0.00, Max: 50.00
    }
    \label{fig:nasa-tlx}
\end{figure*}

\subsubsection{\system{} enables in-context animation comparison}
The participants emphasized that animation comparison and evaluation should occur within the intended environment and camera context, as these factors critically influence judgments of suitability and quality. In general, the participants thought that \system{} supports this in-context comparison (\autoref{fig:usability}b) through camera controls and scene objects. The benefits of such features were highlighted by P9: \textit{“This would give you better intuition, before you actually commit to something. It is more in tune with actual animation workflows than just generate and choose, and then regret”}. Several participants noted that animations evaluated in isolation often fail to account for environmental constraints such as obstacles or scene geometry: \textit{“Some of these motions just seem to ignore the obstacles”} (P10); \textit{“I think a lot happened here when they crossed the first obstacle. A lot of them just go through it”} (P12). This lack of context was seen as a limitation of existing tools, which often present animations in isolation: \textit{“My example is Mixamo, which is completely isolated. It has an animation of a person typing, but it is just them sitting in the air. Having my own context would have been super helpful”} (P1). The participants also highlighted that camera placement and framing can impact the evaluation: \textit{“Especially if I am comparing say which animations are going way above the camera”} (P3). Integrating animations into the scene together with Spatial Lenses allowed users to verify spatial constraints: \textit{“Path was better for those things, like finding the highest point in the animation or when a foot goes through the floor”} (P10). In general, the participants stressed that including environmental and camera context is \textit{“super important and cool”} (P1) to ensure that the animations meet the expected requirements.

\subsection{Tools are Task-dependent}
The participants emphasized that no single visualization was universally best. Instead, the most effective tool depended on the comparison task.  As one participant summarized, \textit{“It boiled down to understanding is it a spatial comparison or is it a temporal comparison? And based on that, I went to the respective Spatial or Temporal Lenses”} (P1). Many participants described switching between different tools as the task changed, noting that each offered unique benefits: \textit{“I switched frequently between Overlay and Grid”} (P3).

\subsubsection{Overlay + Path enables spatial overview in a common frame of reference}
Participants indicated that spatial comparisons, such as identifying which animation extended the farthest, were well supported by \system{} (\autoref{fig:usability}b). They expressed a strong preference for a combination of Overlay and Path for this type of comparison, which was further highlighted in the interaction logs that showed that this combination was used by all participants. This combination allowed participants to make spatial judgments without temporally aligning the animations, while also visualizing them in a common frame of reference to contrast the motions. As one participant explained, \textit{“The Overlay is always useful for comparing space Paths”} (P4). Path was especially valued for revealing movement ranges and were used with joint filtering to focus on specific joints and manage occlusion.

\begin{figure*}
    \centering
    \includegraphics[width=\linewidth]{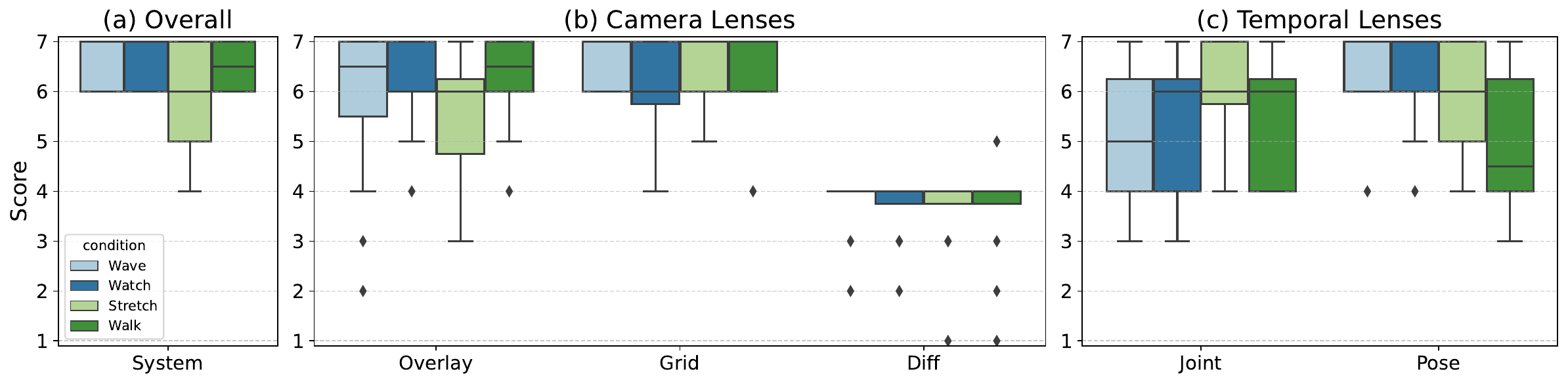}
    \caption{Subjective feedback on perceived usefulness for (a) overall system, (b) Camera Lenses, and (c) Temporal Lenses for each scenario. Results show how exclusion of Root Motion (Wave and Stretch) affects perceived Overlay and Grid usability, longer animations (Wave and Walk) has an effect on perceived usability of the Temporal Lenses.}
    \label{fig:component-ratings}
    \Description{
    Boxplots of usefulness of components and overall system.
(a) Overall:
  System:
    Stretch:
      Mean: 6.00
      Box limits - Q1: 5.00, Median: 6.00, Q3: 7.00
      Whisker limits - Lower: 2.00 (actual: 4.00), Upper: 10.00 (actual: 7.00)
      Data range - Min: 4.00, Max: 7.00
    Walk:
      Mean: 6.50
      Box limits - Q1: 6.00, Median: 6.50, Q3: 7.00
      Whisker limits - Lower: 4.50 (actual: 6.00), Upper: 8.50 (actual: 7.00)
      Data range - Min: 6.00, Max: 7.00
    Watch:
      Mean: 6.58
      Box limits - Q1: 6.00, Median: 7.00, Q3: 7.00
      Whisker limits - Lower: 4.50 (actual: 6.00), Upper: 8.50 (actual: 7.00)
      Data range - Min: 6.00, Max: 7.00
    Wave:
      Mean: 6.58
      Box limits - Q1: 6.00, Median: 7.00, Q3: 7.00
      Whisker limits - Lower: 4.50 (actual: 6.00), Upper: 8.50 (actual: 7.00)
      Data range - Min: 6.00, Max: 7.00

(b) Camera Lenses:
  Overlay:
    Stretch:
      Mean: 5.42
      Box limits - Q1: 4.75, Median: 6.00, Q3: 6.25
      Whisker limits - Lower: 2.50 (actual: 3.00), Upper: 8.50 (actual: 7.00)
      Data range - Min: 3.00, Max: 7.00
    Walk:
      Mean: 6.25
      Box limits - Q1: 6.00, Median: 6.50, Q3: 7.00
      Whisker limits - Lower: 4.50 (actual: 4.50), Upper: 8.50 (actual: 7.00)
      Data range - Min: 4.00, Max: 7.00
    Watch:
      Mean: 6.08
      Box limits - Q1: 6.00, Median: 6.00, Q3: 7.00
      Whisker limits - Lower: 4.50 (actual: 4.50), Upper: 8.50 (actual: 7.00)
      Data range - Min: 4.00, Max: 7.00
    Wave:
      Mean: 5.75
      Box limits - Q1: 5.50, Median: 6.50, Q3: 7.00
      Whisker limits - Lower: 3.25 (actual: 3.25), Upper: 9.25 (actual: 7.00)
      Data range - Min: 2.00, Max: 7.00

  Grid:
    Stretch:
      Mean: 6.50
      Box limits - Q1: 6.00, Median: 7.00, Q3: 7.00
      Whisker limits - Lower: 4.50 (actual: 5.00), Upper: 8.50 (actual: 7.00)
      Data range - Min: 5.00, Max: 7.00
    Walk:
      Mean: 6.50
      Box limits - Q1: 6.00, Median: 7.00, Q3: 7.00
      Whisker limits - Lower: 4.50 (actual: 4.50), Upper: 8.50 (actual: 7.00)
      Data range - Min: 4.00, Max: 7.00
    Watch:
      Mean: 6.00
      Box limits - Q1: 5.75, Median: 6.00, Q3: 7.00
      Whisker limits - Lower: 3.88 (actual: 4.00), Upper: 8.88 (actual: 7.00)
      Data range - Min: 4.00, Max: 7.00
    Wave:
      Mean: 6.67
      Box limits - Q1: 6.00, Median: 7.00, Q3: 7.00
      Whisker limits - Lower: 4.50 (actual: 6.00), Upper: 8.50 (actual: 7.00)
      Data range - Min: 6.00, Max: 7.00

  Diff:
    Stretch:
      Mean: 3.58
      Box limits - Q1: 3.75, Median: 4.00, Q3: 4.00
      Whisker limits - Lower: 3.38 (actual: 3.38), Upper: 4.38 (actual: 4.00)
      Data range - Min: 1.00, Max: 4.00
    Walk:
      Mean: 3.58
      Box limits - Q1: 3.75, Median: 4.00, Q3: 4.00
      Whisker limits - Lower: 3.38 (actual: 3.38), Upper: 4.38 (actual: 4.38)
      Data range - Min: 1.00, Max: 5.00
    Watch:
      Mean: 3.67
      Box limits - Q1: 3.75, Median: 4.00, Q3: 4.00
      Whisker limits - Lower: 3.38 (actual: 3.38), Upper: 4.38 (actual: 4.00)
      Data range - Min: 2.00, Max: 4.00
    Wave:
      Mean: 3.75
      Box limits - Q1: 4.00, Median: 4.00, Q3: 4.00
      Whisker limits - Lower: 4.00 (actual: 4.00), Upper: 4.00 (actual: 4.00)
      Data range - Min: 2.00, Max: 4.00

(c) Temporal Lenses:
  Joint:
    Stretch:
      Mean: 5.92
      Box limits - Q1: 5.75, Median: 6.00, Q3: 7.00
      Whisker limits - Lower: 3.88 (actual: 4.00), Upper: 8.88 (actual: 7.00)
      Data range - Min: 4.00, Max: 7.00
    Walk:
      Mean: 5.50
      Box limits - Q1: 4.00, Median: 6.00, Q3: 6.25
      Whisker limits - Lower: 0.62 (actual: 4.00), Upper: 9.62 (actual: 7.00)
      Data range - Min: 4.00, Max: 7.00
    Watch:
      Mean: 5.42
      Box limits - Q1: 4.00, Median: 6.00, Q3: 6.25
      Whisker limits - Lower: 0.62 (actual: 3.00), Upper: 9.62 (actual: 7.00)
      Data range - Min: 3.00, Max: 7.00
    Wave:
      Mean: 5.17
      Box limits - Q1: 4.00, Median: 5.00, Q3: 6.25
      Whisker limits - Lower: 0.62 (actual: 3.00), Upper: 9.62 (actual: 7.00)
      Data range - Min: 3.00, Max: 7.00

  Pose:
    Stretch:
      Mean: 5.92
      Box limits - Q1: 5.00, Median: 6.00, Q3: 7.00
      Whisker limits - Lower: 2.00 (actual: 4.00), Upper: 10.00 (actual: 7.00)
      Data range - Min: 4.00, Max: 7.00
    Walk:
      Mean: 5.08
      Box limits - Q1: 4.00, Median: 4.50, Q3: 6.25
      Whisker limits - Lower: 0.62 (actual: 3.00), Upper: 9.62 (actual: 7.00)
      Data range - Min: 3.00, Max: 7.00
    Watch:
      Mean: 6.33
      Box limits - Q1: 6.00, Median: 7.00, Q3: 7.00
      Whisker limits - Lower: 4.50 (actual: 4.50), Upper: 8.50 (actual: 7.00)
      Data range - Min: 4.00, Max: 7.00
    Wave:
      Mean: 6.42
      Box limits - Q1: 6.00, Median: 7.00, Q3: 7.00
      Whisker limits - Lower: 4.50 (actual: 4.50), Upper: 8.50 (actual: 7.00)
      Data range - Min: 4.00, Max: 7.00
    }
\end{figure*}

\subsubsection{Grid + Temporal Lenses supported quick timing comparison and visual confirmation}

The participants also indicated that \system{} was useful for temporal comparison (\autoref{fig:usability}b). The interaction logs showed that all participants would rely on the Temporal Lenses' overview of the whole animation (\autoref{sec:time-interval}) to make timing comparisons. Participants would frequently switch between Joint or Pose based on personal preference and the nature of the animation (see \autoref{sec:animation-complexity}), with a slight preference for Pose due to its clear distinction between pose transitions, which was seen as useful for timing comparison as expressed by P3 \textit{“I know where certain poses are starting and where the next pose is starting.”} Temporal Lenses also helped the participants find the relevant timestamp in cases where further visual inspection was needed to confirm the differences. The interaction logs showed that Grid was mainly used (8 participants) for visual inspection of temporal differences  because it allowed participants to see animations side by side for direct timing assessment: \textit{“Grid was better for timing”} (P3). The participants also adjusted the playback speed to slow down fast movements: \textit{“Changing the playback speed was very useful here because things happen really fast”} (P2). In some cases, participants manually aligned the animations to allow direct comparisons, using the information provided by Temporal Lenses as a guide for alignment: \textit{“If they [poses] are not at the same time, the comparison was harder. So you could slide them to align them” }(P6).

\subsubsection{Grid + Animation Filtering/Sequential enables structured visual inspection for action matching and naturalness comparison}

Participants agreed that \system{} supports the inspection of action matching and the comparison of naturalness (\autoref{fig:usability}b). Participants relied on visual inspection of the animations, which led to longer completion times compared to spatial and temporal tasks (\autoref{tab:performance}). Interaction logs showed that participants relied on visual inspection of the animations, using Concurrent playback together with Grid (9 participants) to systematically filter out candidates through the timeline selections of the animations. This process often involved slowing down the playback to ensure accuracy: \textit{“I play it slowly at 0.1 speed because I cannot look at all of them at the same time”} (P2). Some participants relied on Sequential playback (3 participants) combined with Grid or Overlay to make judgments as efficiently as possible for this type of task: \textit{“I will just look at the animation one by one. That is probably the fastest way”} (P6). To speed up the action matching identification process, a subset of participants (5 participants) used the Pose Lens to identify poses that represent the actions and then search for other animations that contained the same pose cluster for further visual inspection.

Participants described naturalness comparisons as inherently subjective and required close visual inspection to assess motion fluidity and realism. The participants particularly expressed a preference for Grid when visually inspecting the naturalness of the animations: \textit{“What I like about Grid is that I can start with all of them and go through”} (P10). In addition, four participants found Trace and Path useful in and amplifying subtle qualities of motion during visual inspection: \textit{“Trace and Path were very cool. They are good for naturalness and smoothness of motion”} (P2).

\subsection{Animation Type Affects Comparison}
In addition to the type of comparison, participants noted that the type of animation used in the different scenarios affected which tools and strategies were more effective. Due to the variety of tools provided by \system{}, the system was perceived as useful (\autoref{fig:component-ratings}a) across all scenarios. This was further highlighted by Friedman tests which did not show significant differences in the perceived usefulness of \system{} between scenarios (\autoref{fig:nasa-tlx}a). Furthermore, Friedman tests of the Raw NASA-TLX results (\autoref{fig:nasa-tlx}b-f) showed significant results for Mental Demand ($\chi^{\scriptscriptstyle 2}(3)$=$12.3$, $p$$<$$.05$, \autoref{fig:nasa-tlx}a) and Effort ($\chi^{\scriptscriptstyle 2}(3)$=$15.0$, $p$$<$$.05$ \autoref{fig:nasa-tlx}c), but post-hoc Bonferroni-corrected Wilcoxon Signed-rank tests showed no significant differences between scenarios. Although we did not find significant differences, trends in usability (\autoref{fig:component-ratings}) and NASA TLX results (\autoref{fig:nasa-tlx}) along with user comments indicated that the Stretch scenario which consisted of a longer animation without root motion was considered more demanding.

\subsubsection{Grid handles occlusion between animations for scenarios without root motion}
Although both the Overlay and Grid were considered useful by the participants, the interaction logs showed that the Overlay was less used for the Wave and Stretch scenarios. Animations without root motion, where characters remained in the same location, often caused severe occlusion in the Overlay view, making visual inspection and comparisons difficult: \textit{“The Overlay view becomes too chaotic because all the arms kind of overlap”} (P2); \textit{“They are all in the same position and there is no horizontal movement. Overlay in this case is really not helpful”} (P7). In contrast, Grid enabled users to see differences between animation side-by-side, avoiding the occlusion issue altogether. This trend can be seen in the participants' reduced perceived usefulness of the Overlay in the Stretch and Wave scenarios (\autoref{fig:component-ratings}b).

\subsubsection{Overlay enables a shared frame of reference in scenarios with root motion}
In contrast, the Walk and Watch scenarios containing root motion introduced a different challenge: characters moved across the environment at varying rates, complicating spatial alignment and side-by-side evaluation. As one participant explained, \textit{“When the animation moves across the camera, it is much harder to compare because they all move at different rates. One character will be on the left [side] and another on the right at different times”} (P2). Overlay was considered more useful in such scenarios as it provided a shared frame of reference for easier alignment and comparison. This trend can also be seen in the higher perceived usefulness of Overlay in the Watch and Walk scenarios (\autoref{fig:component-ratings}b). 

\subsubsection{Longer animations increased complexity} \label{sec:animation-complexity}

For the shorter Watch and Wave scenarios, participants favored Pose Lens due to its simplicity, with Joint Lens only used by a subset of 3 participants.  With extended animation length, Pose Lens became more complex to interpret: \textit{"Pose Lens was still useful, but less so because it got so complex with this movement"} (P1), leading participants to rely more on Joint Lens in the longer Walk and Stretch scenarios. Although no significant differences were found, this trend is reflected in participants' higher perceived usefulness of Pose for shorter scenarios and Joint for longer ones (\autoref{fig:component-ratings}c). Participants also noted that longer animations created larger inconsistencies in where key actions occurred, making alignment more important and timing comparisons difficult: \textit{"I think longer animations are harder to compare because for one of them it could happen at like the start of the animation. For another, a certain movement will happen in the end"} (P2). Finally, as indicated by longer completion times in the Stretch and Walk scenarios (\autoref{tab:study_questions}), extended animations made visual inspection more time-consuming, increasing reliance on playback controls: \textit{“Increasing playback [speed] can save time”} (P12).

\section{Discussion}

\system{} is a system designed to help animators evaluate and compare generated character animations to aid their creative processes when creating animations. Comparison and evaluation of animations are achieved by contextualizing animations in their intended environment (DG1) visualizing whole animations (DG2), multiple modes for comparison (DG3), and information filtering (DG4). Participants with animation experience in our user study highlighted the general benefits and efficacy of our system in comparing animations and how it helps to compare and evaluate different aspects such as spatial, timing and naturalness, and different types of animations in terms of actions, duration, and with or without translational movement. These results indicate \system{}'s flexibility as a tool to support different aspects of 3D character animation evaluation and comparison. This was a key design guideline for \system{} (DG3) by providing alternative techniques to allow flexibility in comparison and workflows. The results of our study also provided several factors that influence the comparison of character animations, which should be considered for future work.

\subsection{Overview-Detail Visualization for Animation Comparison}
The purpose of our study was to uncover how the design and combination of visualizations and tools can best support animators in comparing character animations. Our results clearly showed that visualizations that represent the entire animation timeline provide a critical high-level overview (DG2), enabling users to identify differences without relying on time-consuming playback or manual alignment. This finding aligns with principles from overview-detail patterns~\cite{Min2025OverviewDetail}, which emphasize the importance of a persistent and global context for analytical efficiency. Our results extend these principles to the 3D character animation domain, where complete temporal coverage is not merely beneficial but essential for effective comparison, as also highlighted in other temporal domains~\cite{Huh2025VideoDiff, Wang2024MotionComparator}. 
In \system{} full-time-interval visualizations, such as the Spatial Path Lens and our Temporal Pose and Joint Lenses, allow differences in motion patterns to be perceived at a glance, reducing cognitive load and the need for frame-by-frame inspection. 

In contrast, designs that only show small or time-point intervals of the animations such as Keypose and Skeleton which are frequently used for animation inspection in animation software~\cite{autodeskMaya, blender}, were not considered useful for comparison between animations. These findings suggest that tools should prioritize persistent, full-context representations alongside ephemeral, granular views to support more efficient comparative workflows. 
Such designs would bridge the gap between traditional tools and the demands of multi-animation comparison for more scalable evaluations.

\subsection{Importance of Contextual Comparison}
\system{}'s integration of animations with the additional camera and scene context (DG1) was seen as a key feature in its usability. Environmental constraints, such as spatial boundaries, object interactions, and surface contact, along with camera framing, strongly influence how motion is perceived. Placing animations in context can also save time and effort, as only the portions visible in the final view need to be considered. Our findings reinforce the importance of context-aware evaluation, particularly in pipelines where environmental constraints are strict, such as level design in games or cinematic production. Ideally, comparative analysis should occur with complete environmental and camera information to ensure spatial alignment, maintain visual continuity, and avoid “context-blind” decisions. Tools such as \system{} can address this need by embedding environmental context directly into visualizations or offering seamless import of scene geometry and camera setups to enable accurate and efficient evaluation. 

\hl{Participants further emphasized that such contextual comparison should not exist as a standalone process, but be integrated into existing animation workflows. This reflects a broader need for seamless transitions between tools, where comparison, editing, and evaluation can occur without disrupting established workflows. Prior work on creativity support tools highlights the importance of enabling such “horizontal movement” across tool ecosystems, allowing users to flexibly combine tools rather than being constrained to isolated systems~\cite{Li2023BeyondTheArtifact}. Embedding comparative visualization directly into existing pipelines would therefore improve efficiency, and align with how animators structure their work in practice.}

\hl{Current animation workflows, particularly with generative tools, often require animators to export sequences into external digital content creation tools such as Maya, game engines, or video editing software to evaluate how animations fit within a scene. This separation between generation, comparison, and editing creates fragmented workflows, requiring users to manually move between tools and disrupting iterative exploration.} For generative animation tools, the integration of environmental previews and comparative visualizations early in the design process can enable early exploration to support divergent thinking and exploration~\cite{Suh2024Luminate, Li2023BeyonedTheArtifact}. Such integration would be particularly valuable when combined with further editing capabilities, enabling users to adjust joint positions, fix motion artifacts, or stylize the motion without leaving the tool. Participants emphasized this need, noting the importance of both contextual evaluation and in-place editing to support decision-making and improve motion quality for the future. Embedding these features directly into generative motion pipelines or existing digital content creation tools would create more seamless and efficient workflows, lowering the barrier to adoption and expanding the creative possibilities of AI-assisted animation. Such pipelines would also be useful for other motion generation methods such as motion capture where early in-context integration and comparison of performances can save cost and time.

\subsection{Towards Adaptive and Semantic Animation Comparison}

An additional key insight from our study is that the task of animation comparison is highly situational. Different animation comparison and evaluation goals, such as action matching, spatial alignment, or timing accuracy, require distinct visualization strategies, and no single tool supports all tasks equally well. \system{} supports this by allowing flexible switching between tools at the user's discretion. One future approach is to support adaptive and automatic switching between views or coordinate multiple visualizations so that users can tailor the interface to the question at hand. Presets for tasks such as timing or naturalness comparison could automatically reconfigure views and highlight relevant features. We suggest the following design combinations to be considered:

\begin{description}

\item[Spatial comparison:] Overlay combined with Path enables assessment of movement range and spatial extent  within a common frame of reference without manual alignment. 

\item[Timing comparison:] Grid combined with Temporal Lenses (Pose or Joint) provides a temporal overview for identifying when key poses occur, with side-by-side playback for visual confirmation of timing differences.

\item[Action matching and naturalness:] Grid combined with animation filtering or sequential playback supports systematic visual inspection, allowing users to progressively select candidates. Pose Lens can further accelerate pose matching by locating animations containing similar poses. 

\end{description}

Beyond task type, the properties of the animations also influence which tools are most effective. Animations with large root motion can create severe spatial misalignment over time, while those without root motion may occlude each other and obscure important differences. We suggest the following adaptations based on animation properties:

\begin{description}
\item[Root motion:] Overlay provides a shared frame of reference for animations with root motion, helping users compare characters moving at different rates. Grid is preferable for stationary animations, avoiding the occlusion that arises when multiple characters occupy the same location.

\item[Animation length:] Pose Lens offers a simple temporal overview for shorter animations, while Joint Lens remains readable for longer sequences where increased pose count would otherwise cause clutter. 

\end{description}

We envision future systems could leverage animation or scene metadata to automatically suggest or configure appropriate visualization strategies based on these properties.

\subsection{Future Work and Limitations}
Although participants found \system{} useful for comparative animation analysis, we identify several opportunities for improvement, and future work based on participant feedback. First, in alignment with DG4, the ability to filter or zoom into specific time ranges to reduce clutter and focus on key motion segments would be an interesting venue for future systems, especially for longer animation sequences.

Second, encoding more information into the Spatial and Temporal Lenses could further alleviate the need for visual inspection. Examples could be multi-joint visualizations or descriptive statistics of animations. The Pose Lens was also considered less useful for certain longer animations, due to its inability to handle complex animations with frequent pose switching, cluttering the timeline bar with small segments. To alleviate this issue, Pose Lens could also support different levels of abstraction for switching between visualization of distinct actions, grouped poses, or individual poses to scale with more complex animations. Semantic descriptions of poses and actions through GenAI models could also help with quickly understanding and identify differences between animations.

Third, although the participants thought that \system{} effectively handles in-context comparison, such tasks relied heavily on visual inspection. Similarly, our results showed that tasks such as action matching and naturalness comparison relied primarily on visual inspection of the animations, thus requiring more time. Further Temporal Lenses could be developed that integrate with the surrounding environment, and that summarize or highlight specific issues related to the environment, the wanted action, or animation naturalness. These directions open pathways for developing more flexible, multi-resolution visualization tools that can adapt to comparison needs, and visualization designs that can automatically show the right visualization depending on user needs. Our user study results provide a first step into this by uncovering strategies and efficient tool designs for different animations comparisons.

\hl{We also acknowledge several limitations in our evaluation. First, our study does not include a direct baseline comparison (e.g., sequential playback or toggling), which limits our ability to quantitatively assess performance gains. Instead, our results primarily demonstrate usability and perceived effectiveness of \system{}. Second, the evaluation tasks were designed to isolate specific aspects of animation comparison (e.g., spatial, temporal), which may not fully capture the subjective and holistic criteria, such as weigh, intent, and appeal, that animators emphasize in real workflows. As a result, while the tasks provide controlled insight into how individual features support comparison, they only partially reflect the complex nature of animation, where multiple criteria are considered simultaneously and in context.}

Finally, our study covered a subset of possible animation types. Other animations may require visualizations for easy comparison that is not covered by \system{}. Future work should explore more complex animation types such as multi-character animations, prop usage, and environmental interactions, which may introduce additional layers of comparison complexity. Similarly, our study contained pre-defined environments and camera angles for practical reasons. \system{} and future systems should be extended to include and evaluate different types of camera shots such as moving cameras and multiple cameras to ensure consistency between sequences to support more complete use cases for animators.
\section{Conclusion}


\hl{We presented \system{}, a visual comparison tool for character animations that supports efficient comparison across spatial, temporal, and contextual dimensions.} Through a functional prototype and quantitative and qualitative user feedback, we demonstrated the importance of contextualization, full time-scale visualizations, multi-modal tools, and information filtering to support animation comparison. \hl{Participants expressed interest in integrating \system{} into their existing workflows to support more efficient evaluation and iteration of animation variants. While recent advances in generative tools make it easier to produce multiple animation alternatives, the need for effective comparison spans traditional, motion capture, and generative pipelines alike. We see \system{} as a step toward tools that better support how animators explore, evaluate, and refine motion, enabling more efficient and informed decision-making within existing creative workflows.}

\begin{acks}
Special thanks to Tovi Grossman and Justin Matejka for their help with ideation, Angie Foss for participant recruitment, and Mithila Majithia for the video figure voiceover. The prototype was implemented with assistance from Cursor and Anthropic Claude, with all generated code reviewed, edited, and tested by the authors.
\end{acks}

\balance
\bibliographystyle{ACM-Reference-Format}
\bibliography{references}

\end{document}